\documentclass{aa}  
\usepackage{xspace}
\usepackage{graphicx}
\usepackage[varg]{txfonts}

 \usepackage{hyperref}
 
 \newcommand{\ATLAS}{{\bf ATLAS}\xspace} 
 \def\bk{\mbox{\boldmath$k$}}
 \def\bq{\mbox{\boldmath$q$}}
 \newcommand{\qp}{q^\prime}
 \newcommand{\nr}{n_{\mathrm{r}}}
 \newcommand{\mps}{m s$^{-1}$}

\begin{document} 

\title{An improved multi-ridge fitting method for ring-diagram helioseismic analysis}


\author{Kaori Nagashima\inst{1}
          \and
          Aaron C. Birch\inst{1}
          \and
          Jesper Schou\inst{1}
          \and 
          Bradley W. Hindman\inst{2}
          \and 
          Laurent Gizon\inst{1,3}
          }

   \institute{
Max-Planck-Institut f\"ur Sonnensystemforschung, 
 Justus-von-Liebig-Weg 3, 37077 G\"ottingen, Germany\\
\email{nagashima@mps.mpg.de}
         \and
  JILA \& Department of Applied Mathematics, University of Colorado, Boulder, CO 80309-0440, USA       
         \and
  Institut f\"ur Astrophysik, Georg-August-Universit\"at G\"ottingen, 
  Friedrich-Hund-Platz 1, 37077 G\"ottingen, Germany
             }

   \date{Received September 10, 2019; accepted November 6, 2019}

\abstract
   {There is a wide discrepancy in current estimates of the strength of convection flows in the solar interior
   obtained using different helioseismic methods applied to observations from the Helioseismic and Magnetic Imager (HMI) onboard the Solar Dynamics Observatory (SDO).  The cause for these disparities is not known.}
   {As one step in the effort to resolve this discrepancy, we aim to characterize the multi-ridge fitting code for ring-diagram helioseismic analysis that is used
   to obtain flow estimates from local power spectra of solar oscillations.}
   {We updated the multi-ridge fitting code developed by Greer et al. (2014, Sol. Phys., 289, 2823)
   to solve several problems we identified through our inspection of the code.  In particular, we changed the 1) merit function to account for the smoothing of the power spectra, 2)~ model for the power spectrum, and 3)~ noise estimates.
   We used Monte Carlo simulations to generate synthetic data and to
   characterize the noise and bias of the updated code by fitting these synthetic data. }
   {The bias in the output fit parameters, apart from the parameter describing the amplitude of the p-mode resonances in the power spectrum, is
   below what can be measured from the Monte-Carlo simulations.  The amplitude parameters are underestimated; this is a consequence
   of choosing to fit the logarithm of the averaged power.  We defer fixing this problem as it is well understood and not significant for measuring flows in the
   solar interior.  The scatter in the fit parameters from the Monte-Carlo simulations is well-modeled by the formal error estimates from the code.
   }
   {We document and demonstrate a reliable multi-ridge fitting method for ring-diagram analysis. The differences between the updated fitting results and the original results are less than one order of magnitude and therefore we suspect that the changes will not eliminate the aforementioned orders-of-magnitude discrepancy in the amplitude of convective flows in the solar interior.
   }

\keywords{Sun: helioseismology --
                Methods: data analysis 
               }

\titlerunning{An improved multi-ridge fitting method }
\authorrunning{K. Nagashima et al.}

\maketitle
%

\section{Introduction}\label{sec:intro}

Understanding solar interior dynamics is crucial to understanding the mechanisms of the solar dynamo.  As one example, convection may play an important role in the formation of magnetic flux tubes, 
as well as in their rise through the convection zone
and their tilts at the solar surface \citep[e.g.,][]{Brun2017}.
Helioseismology, which uses observation of oscillations on the solar surface, is an important probe of interior dynamics.   

Currently there is a major discrepancy between the time-distance \citep{HDS2012} and ring-diagram \citep{2015ApJ...803L..17G} estimates of the strength of
solar subsurface convection at large spatial scales.  
The measurements from time-distance helioseismology
suggest flows orders-of-magnitude weaker than those seen in convection simulations (e.g., in 
the anelastic spherical harmonic (ASH) convection simulations by
\citealt{2008ApJ...673..557M}), while
the measurements from ring-diagrams are closer to the expectations from simulations.

Time-distance helioseismology \citep{1993Natur.362..430D, 1997ASSL..225..241K} 
is based on measuring and interpreting the travel times
of acoustic and surface-gravity wave packets.  These travel times are measured from the temporal cross-covariances between the Doppler observations at pairs of points on the solar surface 
\citep[see][for a review]{Gizon2005}. 
Ring-diagram analysis \citep{1988ApJ...333..996H, 2007AN....328..257A}
measures the Doppler shift of acoustic and surface-gravity oscillation modes in the local power spectra  and uses these Doppler shifts to infer local flows in the solar interior.
To compute the local power spectra, the solar surface is divided into a number of spatial tiles. Each tile is tracked at a rate close to the solar rotation rate. The power spectra of the solar oscillations (Doppler observations) are computed for each tile. 
In the three-dimensional power spectra, roughly concentric rings with high power are present at each frequency and they correspond to the modes of different radial orders.
Flow and wave-speed anomalies in the Sun shift and distort the rings and, hence, one can obtain information about the solar interior from the ring parameters.  
Flow maps from Doppler observations by the Helioseismic and Magnetic Imager \citep[HMI;][]{2012SoPh..275..229S} onboard the Solar Dynamics Observatory \citep[SDO;][]{2012SoPh..275....3P} are automatically computed by the SDO/HMI ring-diagram pipeline \citep{2011JPhCS.271a2008B, 2011JPhCS.271a2009B} on a daily basis since the SDO launch in 2010.

The HMI ring pipeline codes separately fit each single ridge (single radial order $n$) in the power in slices at constant horizontal wavenumber or slices in temporal frequency. \cite{2014SoPh..289.2823G} developed
an alternative approach based on simultaneously fitting multiple ridges (multiple radial orders) at each horizontal wavenumber. 
\cite{2015ApJ...803L..17G} introduced another innovation: they chose a much denser tile layout than other ring-diagram analyses: 16-degree tiles with the separation of 0.25~degree instead of the 7.5~degree spacing (at the equator, the spacing increases at higher latitude to maintain 50\% overlap between neighboring tiles) used for 15-degree tiles in the HMI ring pipeline.  
As described in detail in Appendix~\ref{apsec:original_model},
the \ATLAS code from \cite{2014SoPh..289.2823G, 2015ApJ...803L..17G} provides seven parameters for each ridge and five parameters for the background power at each wavenumber $k$.
\citet{2015ApJ...803L..17G} applied a three-dimensional flow inversion 
to these fit results to estimate the three-dimensional flow field in the solar interior.  Both the dense packing of tiles and the three-dimensional inversions are unique to \ATLAS, however, in this paper we focus only on the fitting component of the code.

As one step toward understanding the causes of the above-mentioned disagreement 
between the helioseismic measurements of subsurface convection,
here we focus on the ring-diagram analysis described by \cite{2015ApJ...803L..17G}.
We revisit the analysis code \citep{2014SoPh..289.2823G, 2015PhDT.......167G} that
was used in that work and identify several issues through a step-by-step examination of the code.
In response to these issues, we have developed an updated method 
and characterize the updated code by applying it to
synthetic data generated from Monte-Carlo simulations.

\section{Description of the updated code}

In this section, we describe our updated \ATLAS code.
Each of the updates is a response to a problem that we found
in our inspection of the original code.  In this
section we will refer to Appendix~\ref{apsec:original_code} for the details
of the original code.

After computing the power spectrum for a particular tile,
the processing steps are:
1) remap the power spectrum from Cartesian $(k_x,k_y)$ to polar $(k,\theta)$ coordinates,
2) rebin the power in azimuth $\theta$; the number of grid points in $\theta$ is reduced from $n_{\mathrm{pix}}=256$ to $n_{\mathrm{pix}}=64$,  
3) fit the logarithm of a Lorentzian model to the logarithm of the smoothed power by least-squares minimization at each $k$ using the Levenberg-Marquardt \citep{1963Marquardt} technique; the model function has $7n_{\mathrm{r}}+ 4$ parameters at each horizontal wavenumber $k$, where $n_\mathrm{r}$ is the number of ridges at the particular value of $k$, and 
4) estimate the covariance matrix of the errors of the fitted parameters 
by computing the inverse of the Hessian matrix of the cost function.
In the following subsections we describe the changes that we introduced
to each of these steps.

 \subsection{Re-binning in azimuth} \label{sec:alt_rebin}

 The computed local power spectra $O(k_x, k_y, \nu)$ and the interpolated spectra
 in polar coordinates are non-negative
 by construction.  
 Following the original code, the interpolated spectra have 256 pixels at each $k$. At $k R_\odot \equiv \ell \sim 500$ (where $R_\odot$ is the solar radius), which corresponds to $k_\mathrm{pix} \equiv k/h_k = 21$ (where $h_k=3.37 \times 10^{-2} \mathrm{Mm}^{-1}$ is the grid spacing in $k$) and which we use in most of the Monte-Carlo test calculations shown later,  this is about twice of the number of grid points in $\theta$ from the full resolution, which has $n_{\mathrm{pix}} \approx 2 \pi k/ h_k$.  
 For the sake of computational efficiency, 
 it is desirable to reduce the number of points in azimuth.
 In the updated code, we use a running box-car smoothing of four-pixel width followed by subsampling by a factor of four to reduce to the number of grid points in $\theta$. 
 This procedure ensures that the resulting
 smoothed power spectrum $O_{\rm s}(k, \theta, \nu)$ will be positive.
We expect that as flows produce $\theta$ variations that are dominantly at azimuthal wavenumber one, it should be possible retain only 
 very low resolution in $\theta$; Sect.~\ref{sec:dis_rebin} discusses this in more detail.
 We retain 64 points in $\theta$ for the examples shown in this paper. 
 
 The original code used a low-pass Fourier filter to smooth the
power spectrum in the $\theta$ direction.
This procedure produces occasional points where the smoothed power is negative.

\subsection{Least-squares fitting of the logarithm of power}\label{subsec:LSfitting}

In the updated code, we use a least-squares fitting
to fit the logarithm of the model power to
the logarithm of the remapped and smoothed power.
Following  \cite{2014SoPh..289.2823G},
the fitting is carried out independently at each horizontal wavenumber $k$. 
As the amount of smoothing
is increased, the probability distribution function (PDF) of the power, as well as its logarithm, approaches a normal distribution (see Appendix \ref{apsec:dist_func_comp} for more details), and it is therefore appropriate to use a least-squares fit .
The logarithm of the smoothed power has the convenient property that the variance of the logarithm of power ($\sigma_N$ in Eq.~(\ref{eq:logP_pdf})) depends only on the details of the remapping and smoothing 
and does not depend on $\theta$ or $\nu$ (see Appendix~\ref{apsec:pdf_logP} for details).

In our approach, the cost function at a single $k$ is:
\begin{equation}
-\ln L(\bq) =   \sum_j \left[\ln{O_s(\theta_j, \nu_j)} - \ln{P(\theta_j, \nu_j; \bq)}\right]^2 \ ,  \label{eq:-lnL2}
\end{equation}
where  $O_{\rm s}(\theta,\nu)$ is the observed spectrum at some fixed $k$ after smoothing in $\theta$
and $P(\theta, \nu; \bq)$ is the model of the spectrum with model parameters $\bq$. 
The summation is taken over all bins $j$ within the fitting range.  
We note that as the error estimates for $\ln{O_{\rm s}}$ ($\sigma_N$ in Eq.~(\ref{eq:logP_pdf})) are all the same, we  set them all to one in writing the cost function. 
For the sake of readability, throughout the remainder of this paper we do not introduce or carry notation to denote the value of $k$; the fitting problems at each $k$ are treated as independent and we will not be comparing fit parameters for different values of $k$.
This is an approximation as the interpolation from $(k_x,k_y)$ space to $(k,\theta)$ space does imply error correlation between the fit parameters at different values of $k$.

In the updated code, we use the Levenberg-Marquardt technique to solve the 
minimization problem.  In particular, we use mpfit.c \citep{2009ASPC..411..251M}, one of the codes in the MINPACK-1 least-squares fitting library \citep{1978More, 1993More}. The covariance matrix of the errors associated with the fitted parameters are estimated at the last step of the Levenberg-Marquardt procedure.  
As a practical note, the error estimates obtained by this method must be scaled as we have assumed $\sigma=1$ in Eq.~(\ref{eq:-lnL2}); see original PDF, Eq.~(\ref{eq:logP_pdf}). In general, calling the code with
an incorrect estimate of $\sigma$ could cause poor performance of the fitting algorithm.  In the current case $\sigma$ is not far from one ($\sigma \sim 0.5$, see Sect.~\ref{subsec:error_estimate}) and we do not expect that this is a significant issue here.  To implement the $\sigma$ estimation in the code is a task for the future.

The original \ATLAS code used 
a maximum-likelihood method based on the assumption that the power spectrum in a single bin in $(k_x, k_y, \nu)$ space follows a chi-squared distribution with two degrees of freedom; see Appendix~\ref{apsec:pdf_raw} for the original likelihood function.  This method does not account for the impact of smoothing on the PDF of the power spectrum.

\subsection{The functional form of the power spectrum} \label{sec:alt_model}

The model function for the power spectrum at a single $k$ in our updated code is
 \begin{eqnarray}
 && P (\theta, \nu; \bq_{\mathrm{BG}}) =\nonumber  \\ 
&& \sum^{n_{\mathrm{r}}-1}_{n=0} \frac{q_{1,n} (q_{2,n}/2) F(\theta;q_{5,n} , q_{6,n}) }{[\nu - q_{0,n} +
k (q_{3,n} \cos \theta + q_{4,n} \sin \theta) /(2\pi)]^2+(q_{2,n}/2)^2} \nonumber \\
&& + B(\theta, \nu; \bq_{\mathrm{BG}}) \ ,
 \label{eq:altered_Pmodel}
 \end{eqnarray}
 where 
 \begin{equation}
  F(\theta; p_1, p_2) = 1+ p_1 \cos{(2 \theta)}+ p_2 \sin{(2\theta)} \;  .  \label{eq:altered_Pmodel_anis}
   \end{equation}
The fitting parameters for the peaks ($n=0, \dots n_{\mathrm{r}}-1$, where $n_{\mathrm{r}}$ is the number of the ridges) are $q_{0,n}=\nu_n$ is the frequency of the $n$-th peak, $q_{1,n}=A_n$  is the amplitude, $q_{2,n}=\Gamma_n$ is the width, 
$(q_{3,n},q_{4,n})=\vec{u} = (u_{x,n},u_{y,n})$ is the horizontal velocity, $q_{5,n}=f_{c,n}$ and $q_{6,n}=f_{s,n}$ are parameters to handle anisotropy. 
The background is modeled at each $k$ as 
\begin{equation}
B(\theta,\nu; \bq_{\mathrm{BG}}) =\frac{{q}_{0,\mathrm{BG}}}{\nu^{q_{1,\mathrm{BG}}}}  F(\theta; q_{2,\mathrm{BG}}, q_{3,\mathrm{BG}}) \ ,  \label{eq:altered_Bmodel}
\end{equation}
where 
$F$ is again given by Eq.~(\ref{eq:altered_Pmodel_anis})  
and 
$q_{0,\mathrm{BG}} = B_0$ is the amplitude,  
$q_{1,\mathrm{BG}}=b$ is the power-law index,
$q_{2,\mathrm{BG}}=f_{c, \mathrm{bg}}$ and $q_{3,\mathrm{BG}}=f_{s,\mathrm{bg}}$ are the parameters to handle the anisotropy.
The number of the parameters in total is $7 n_{\mathrm{r}}+ 4$ at each $k$.
For reference, Tables~\ref{table:input_parameters_k14}, \ref{table:input_parameters_k21}, and \ref{table:input_parameters_k42}
show the physical meaning of each of the fitting parameters.

We altered the model function Eq.~(\ref{eq:altered_Pmodel}) 
from the original Eq.~(\ref{eq:Pmodel})
to make the parameterization more stable.
The following subsections describe the motivation for these changes.

\subsubsection{Parameters of the anisotropy terms}\label{sec:alt_anis}

We replaced $F^\prime$ defined by Eq.~(\ref{eq:Pmodel_anis}) with $F$ defined by Eq.~(\ref{eq:altered_Pmodel_anis})
and used the same form for the anisotropy terms for the background,
although the exact form for the background function is subject to other alterations discussed in Sect.~\ref{sec:alt_backparams}.

Our alteration does not change the space of functions that can be fit with $F(\theta)$ but it does not suffer from the issue of indeterminate phase for nearly isotropic power spectra.
In the original parameterization, the amplitudes of the anisotropic part of the model function  ($\qp_{5,n}$ and $\qp_{3,\mathrm{BG}}$) 
 are usually much smaller than one.
 As a consequence, 
 the phase ($\qp_{6,n}$ or $\qp_{4,\mathrm{BG}}$) does not matter much in the fitting, and, hence, it is unstable. 
In the particular case of isotropic power spectra with $\qp_{5,n}=0$ for all $n$ and $\qp_{3,\mathrm{BG}}=0$, the phases $\qp_{6,n}$ and $\qp_{4,\mathrm{BG}}$ are indeterminate.

\subsubsection{Parameters of the background model} \label{sec:alt_backparams}

We also changed the background parameterizations for the sake of stability.
The background model (Eq.~(\ref{eq:Bmodel})) originally contained five parameters.
In this section we explain our motivation for reducing this to the four parameters shown in Eq.~(\ref{eq:altered_Bmodel}). 

The background model in Eq.~(\ref{eq:Bmodel}) is based on the model of \cite{1985ESASP.235..199H}:
\begin{eqnarray}
B_\mathrm{Harvey}(\nu) = \frac{4 \sigma_{\mathrm{rms}}^2 \tau}{1+(2\pi \nu \tau)^2} \ ,
\end{eqnarray}
where $\tau$ is the characteristic timescale of the velocity field in question and $\sigma_{\mathrm{rms}}$ is the rms velocity.
In this case the index in the original background model (Eq.~(\ref{eq:Bmodel})) would be $q_{2,\mathrm{BG}}^\prime=2$.
\citet{2002ESASP.508...47A} suggested a generalization of the \cite{1985ESASP.235..199H} model,
where $q_{2,\mathrm{BG}}^\prime$ can be the range of 2 to 6.

In the \ATLAS fittings,  however,
we typically obtain the index $q_{2,\mathrm{BG}}^\prime\sim 1$ for HMI observations. 
Also, the roll-off frequency obtained from the fitting of HMI observations, $q_{1,\mathrm{BG}}^\prime$, is quite low ($\sim 1$~$\mu$Hz) and below the frequency resolution for 28.8-hour power spectra (9.7~$\mu$Hz), which 
is the typical observation length for 15-degree tiles in HMI pipeline and 16-degree tiles in \ATLAS. 

This suggests that the background is not related to either the
supergranulation ($\tau \approx 10^5$~sec, or $\nu \approx 10$~$\mu$Hz) or granulation time scales ($\tau \approx 4\times 10^2$~sec, $\nu \approx 2.5$~mHz),
based on Table~1 of  \cite{1985ESASP.235..199H}. 
As these background parameters are not consistent with the original physical model, we might need to reconsider the background model. At this moment, however, this is a task for the future, and 
we retain this model in the altered form mentioned below. 

In the case of $\nu \gg \qp_{1,\textrm{BG}}$ , the original background model, Eq.~(\ref{eq:Bmodel}), can be simplified:
\begin{eqnarray}
&&\frac{\qp_{0,\textrm{BG}}}{1+(\nu/\qp_{1,\textrm{BG}})^{\qp_{2,\textrm{BG}}}} F^\prime (\theta; \qp_{3,\textrm{BG}}, \qp_{4,\textrm{BG}}) \nonumber \\
&&\sim \frac{\qp_{0,\textrm{BG}} \  {\qp_{1,\textrm{BG}}}^{\qp_{2,\textrm{BG}}}}{\nu^{\qp_{2,\textrm{BG}}}} F^\prime(\theta; \qp_{3,\textrm{BG}}, \qp_{4,\textrm{BG}}) , 
\end{eqnarray}
 we therefore redefine the background model $B(\theta, \nu)$ with four parameters $q_{i,\textrm{BG}} \ (i=0,1,2,3)$ with the altered anisotropy terms in the form of 
 Eq.~(\ref{eq:altered_Bmodel}). 
 As the new background model ($q_{i,\textrm{BG}}$) has only four parameters instead of the original five,
 the index $i$ has different meanings in the two models.

\section{Performance of the updated code} \label{sec:performance}
 
Monte-Carlo simulations are a powerful tool for testing fitting methods.
In this section, we use this approach to characterize the performance
of the updated fitting code.
In Sect.~\ref{subsec:error_estimate} we compare 
the scatter of the fitting results of Monte-Carlo simulations with the noise estimated by the updated code.  In Sect.~\ref{subsec:flow_singlepeak}
we measure the bias of the flow estimates for some simple cases.

\subsection{Error estimates} \label{subsec:error_estimate}

We use the approach of \cite{2004ApJ...614..472G} to generate realizations
of the wavefield.  The assumption of this approach is that
the real and imaginary parts of the wavefield at each point $(k_x, k_y, \nu)$
are independent Gaussian random variables.
In more detail, the method is as follows:
1) create a Lorentzian model (Eq.~(\ref{eq:altered_Pmodel})) for the limit spectrum using a set of input parameters,
2) pick two standard normally distributed random numbers at each grid point $(k_x, k_y, \nu)$ and take sum of the squared numbers divided by two to make a chi-square distribution with two degrees of freedom, and
3) multiply the limit spectrum by these random numbers to obtain one realization of the power spectrum.   

After each realization of the power spectrum is generated in $(k_x, k_y, \nu)$ space, it is then remapped and rebinned in the same manner as the local power spectra computed
from the observations.
To obtain the parameters for the input model power spectrum in the first step above, 
we used the average over Carrington rotation 2211 of the power spectra for the disc center tile
from the SDO/HMI pipeline (these average power spectra were obtained using the Data Record Management System (DRMS) specification {\tt hmi.rdvavgpspec\_fd15[2211][0][0]}). 
We then used the updated code to fit these average power spectra. 
The parameters resulting from these fits at 
$\ell =328$, $492$, and $984$ ($k_\mathrm{pix}=14$, $21$, and $42$, respectively) 
are shown in Tables~\ref{table:input_parameters_k14}, 
\ref{table:input_parameters_k21}, and \ref{table:input_parameters_k42}, respectively,
in Appendix \ref{apsec:inputparams}.
We later use these parameters to generate Monte-Carlo synthetic data.

\begin{figure}[hb]
\centering
\includegraphics[width=0.48\textwidth]{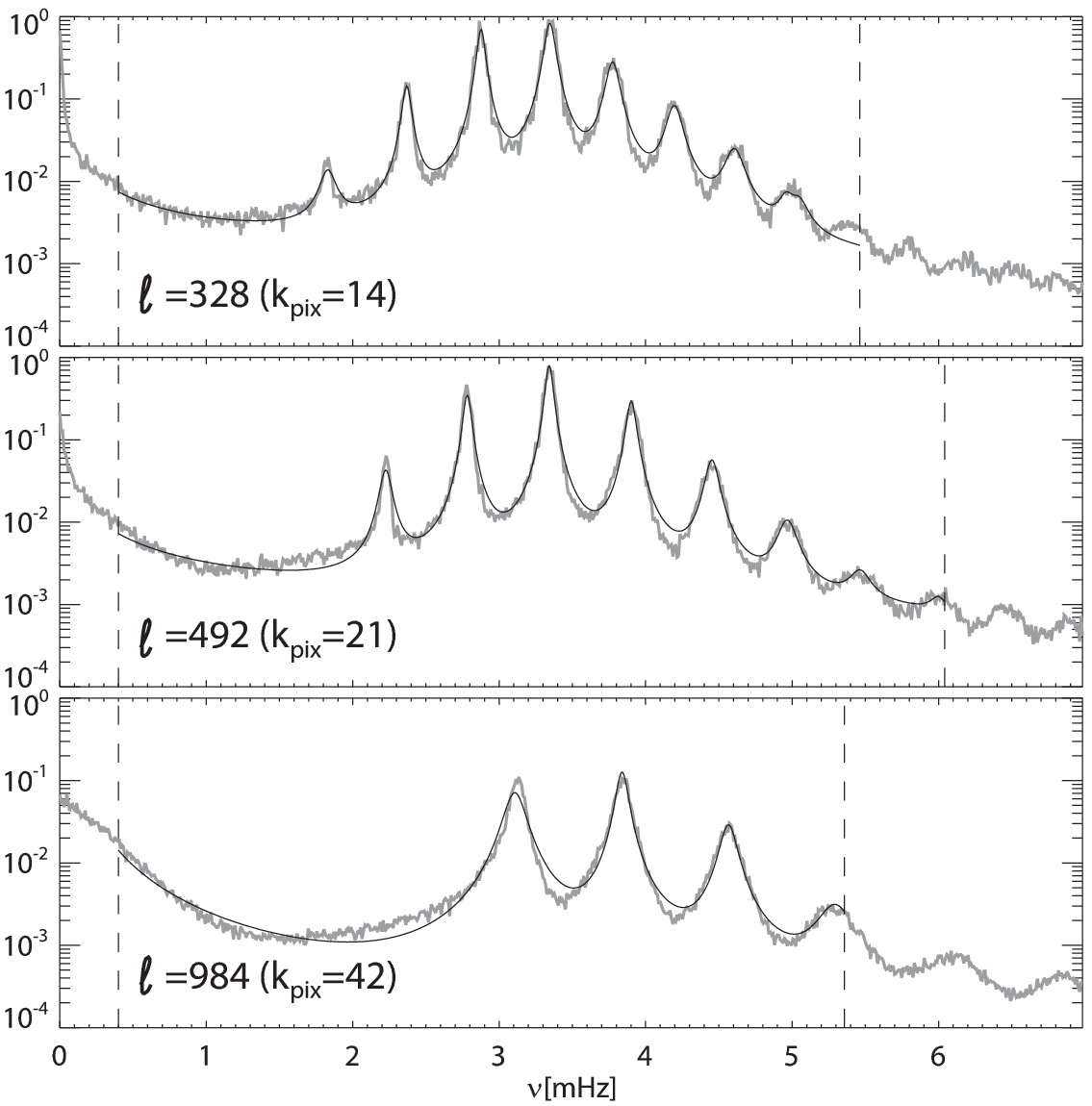}
\caption{Slices through the input limit spectrum  (black) and 
 observational spectrum  averaged over one Carrington rotation (gray), which was 
used to obtain the input parameters, 
at $\theta=0$ and a few different values of $k$.   
Vertical dashed lines indicate the fitting ranges and the models are plotted only within these ranges.
Lower limits for all $k$ are fixed at 0.4 mHz, 
while the upper limits depend on the initial guess for each $k$, and they are 
the highest peak frequencies plus the widths of the the peak.  
For the cases of $\ell=328$ and $492$ ($k_\mathrm{pix}=14$ and $21$, respectively), the peaks with the radial order, $n$, from 0 to 7 are used in the fitting,
while for the case of $\ell=984$ ($k_\mathrm{pix}=42$), $0\le n \le 3$. 
}
\label{fig:modelfunc}
\end{figure}

Figure~\ref{fig:modelfunc}  shows a few slices of the input limit spectrum, namely, Eq.~(\ref{eq:altered_Pmodel}) evaluated for the parameters ${\bf q}$ listed in 
Tables in Appendix \ref{apsec:inputparams}.
The original observed power spectrum is also shown.
Figure~\ref{fig:modelfunc} shows that the model power spectrum is reasonable compared to the observables,
although the details of the peak shapes and the background slopes are not always reproduced by the model and there is still room for improvement with regard to the model function.

We created 500 realizations of the power spectrum using the above procedure. 
 The fitting code occasionally produces outliers and for these computations, we removed them.  
Specifically, we removed outliers iteratively: in each iteration we computed the standard deviation of the samples between the tenth and 90th percentile and we discarded points at more than five times the standard deviation away from the mean.  We repeated this procedure until no further points were removed.
After this outlier removal, the number of valid samples are 489 ($\ell=328$, $k_{\mathrm{pix}}=14$), 500 ($\ell=492$, $k_{\mathrm{pix}}=21$), 
and 496 ($\ell=984$, $k_{\mathrm{pix}}=42$) out of 500 samples.

\begin{figure*}[hbp]
\includegraphics[width=0.9\textwidth]{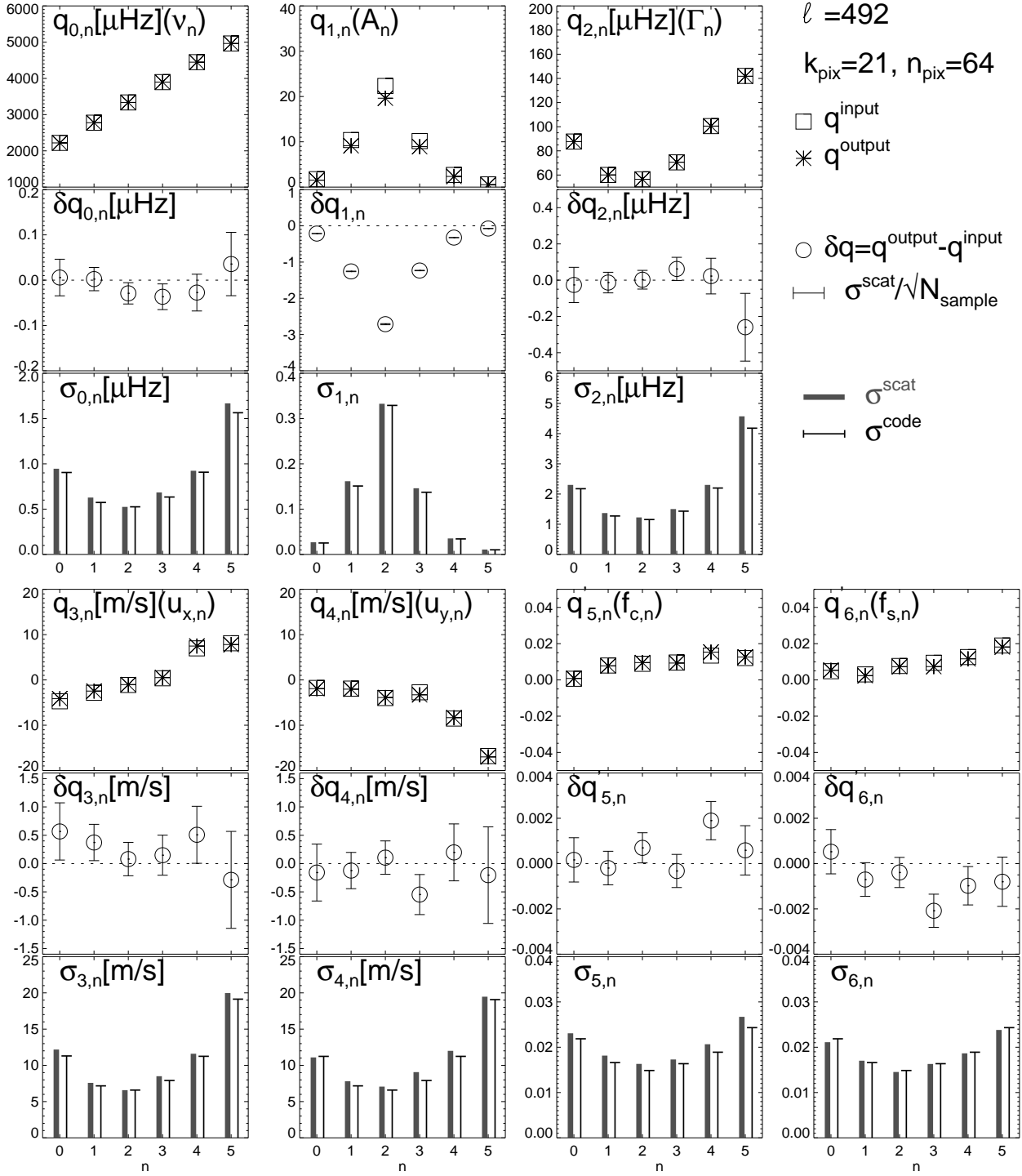}
\caption{Fitting parameters and their errors for the case $\ell = 492$  ($k_{\mathrm{pix}}=21$).
Each panel has three parts.  
In the upper part of each panel, squares indicate the input parameters, $q_{i,n}^{\mathrm{input}}$.
 The asterisks indicate the fitting results obtained by the updated code, namely, the mean of Monte-Carlo samples, $q_{i,n}^\mathrm{output}$. 
In the middle part of each panel, the deviation of the mean fitting results from the input, $\delta q_{i,n} = q_{i,n}^{\mathrm{output}} - q_{i,n}^{\mathrm{input}}$, are illustrated by the circles and the expected scatter of the mean  computed from the square root of variance $\sigma^\mathrm{scat}$ of the 500 Monte Carlo samples divided by the square root of the sample number, $N_\mathrm{sample}$ (here it is 500) are shown as the error bars.
The dashed horizontal lines are at $\delta q = 0$. 
In the lower part of each panel the scatter of the 500 samples, $\sigma^{\mathrm{scat}}$ and the scaled error estimated by the updated code, $\sigma^\mathrm{code}$, are depicted by the thick gray lines and the thin lines with short horizontal bars on the edge.
Errors estimated by the updated code are scaled by $\sigma(\alpha=1.5, k_{\mathrm{pix}}, n_{\mathrm{pix}})=0.569$ as described in Appendix~\ref{apsec:pdf_logP}.
The error bars on the middle panel of $\delta q_{1,n}$ are tiny at this scale; 
the underestimation is relatively large. 
However, the errors of $\delta q_{1,n}$ are $\sigma^\mathrm{scat}/\sqrt{N_\mathrm{sample}}$ and, therefore, available from $\sigma^\mathrm{scat}$ in the lower panel 
and $N_{\mathrm{sample}}=500$; they are $~\sim 0.02$ at most. 
}
\label{fig:kr21_p_dp}
\end{figure*}

Figure~\ref{fig:kr21_p_dp} shows the average and scatter of the output peak parameters associated with the $n=0, \dots 5$ ridges for $\ell=492$ ($k_{\rm pix}=21$)  from fitting 500 Monte-Carlo realizations of the power spectrum.
Table~\ref{table:Lor_model_background} shows the average and scatter of the output background parameters 
from the same set of fitting results.
From Fig.~\ref{fig:kr21_p_dp} and Table~\ref{table:Lor_model_background}, we see that most of the averages of the parameters estimated by the fitting are within the expected scatter in the mean ($\sigma^\mathrm{scat}/\sqrt{N_\mathrm{sample}}$, where $N_\mathrm{sample}$ is the number of Monte-Carlo realizations, hence $N_\mathrm{sample}=500$ here).  The amplitude is an exception; it is always smaller than the input. 
This is the result of taking the logarithm of the smoothed power; see Appendix~\ref{apsec:app_amp} for details. 
Also, from Fig.~\ref{fig:kr21_p_dp} and Table~\ref{table:Lor_model_background}, we see that the error estimated by the updated code $\sigma^{\mathrm{code}}$ is consistent with the scatter of the samples $\sigma^{\mathrm{scat}}$.  This shows that the error estimates produced by the updated code are reasonable.

We carried out the analogous Monte-Carlo simulations for the cases of twice larger $k$ and two-thirds $k$.
Although the number of peaks to be fitted is not the same in these cases as the number of prominent peaks is smaller for larger $k$, the behavior of the fitting results and their error estimates are similar to those in the case shown in Fig.~\ref{fig:kr21_p_dp} and Table \ref{table:Lor_model_background}.

\begin{table*}[hbp]
\caption{
Background parameters in the input model for the Monte Carlo simulation and the fitting results at $\ell = 492$ ($k_{\mathrm{pix}}=21$). 500 realizations were used. Figure~\ref{fig:kr21_p_dp} shows the  corresponding fitting results for the peak parameters. The standard deviations over the 500 realizations and the scaled errors provided by the updated code are consistent.}
\label{table:Lor_model_background}    
\centering                                      
\begin{tabular}{c | c c c c}          
\hline\hline                        
\centering
parameters& input &\multicolumn{3}{c}{fitting results (500 samples)} \\
 \cline{3-5} 
 && mean & standard deviation & scaled error by code \\
\hline
$q_{0,\textrm{BG}}$  ($B_0$) & $  1.959$ &$  1.738$ &$  0.191$ &$  0.180$\\
$q_{1,\textrm{BG}}$  ($b$) & $  0.947$ &$  0.948$ &$  0.016$ &$  0.015$\\
$q_{2,\textrm{BG}} [10^{-2}]$  ($f_{c, \mathrm{bg}}$)& $  4.484$ &$  4.612$ &$  1.020$ &$  0.952$\\
$q_{3,\textrm{BG}} [10^{-2}]$ ($f_{s, \mathrm{bg}}$) & $ -0.059$ &$ -0.077$ &$  0.955$ &$  0.953$\\
\hline                                             
\end{tabular}
\end{table*}

\subsection{Bias of the flow estimates and correlations between the fitting parameters} \label{subsec:flow_singlepeak}

To  measure the bias of the flow estimates and the correlations between the fitting parameters,  
we make models with a simple flow:  isotropic models, except for a specific flow $u_x$ for the $n=3$ ridge.

The parameters for the model are identical to those from Sect.~\ref{subsec:error_estimate},
except for the parameters related to the azimuthal angle $\theta$; namely, the peak parameters $q_{3,n}$, $q_{4,n}$, $q_{5,n}$, and $q_{6,n}$ (for all $n$) and the
background parameters $q_{2,b}$ and $q_{3,b}$ are all set to zero.
The flow $u_{x,n}$ ($q_{3,n}$) is non-zero for a single $n$, $n=3$.
The limit spectrum is constructed with Eq.~(\ref{eq:altered_Pmodel}) and these input parameters, and
the realizations of the power spectrum are generated
in the same way as those in Sect.~\ref{subsec:error_estimate}. 

In this subsection we compare results from the updated code with results from the original code and also a modified version of the original code.
While the  original code fits a Lorentzian model to the square root of the power, the  modified original code fits a Lorentzian model to the power itself. 
Although the original and modified original codes assumes the five-parameter background model (Eq.~(\ref{eq:Bmodel})), 
here we created the limit spectrum using the four-parameter background model (Eq.~(\ref{eq:altered_Bmodel})),
which we use in our updated code.
As described in Sect.~\ref{sec:alt_backparams},  under the current conditions, 
our updated four-parameter background model can approximate well the original five-parameter background model and the original and the modified original codes are applicable, although we need to keep in mind that the 
meaning of the background parameters in the results from different codes are different. 
Comparable tests using input parameters obtained by the original code
give similar correlation coefficients maps. 

\subsubsection{Bias of the flow estimate}\label{subsec:biasofflow}

\begin{figure*}[hpbt]
\centering
\includegraphics[width=0.75\textwidth]{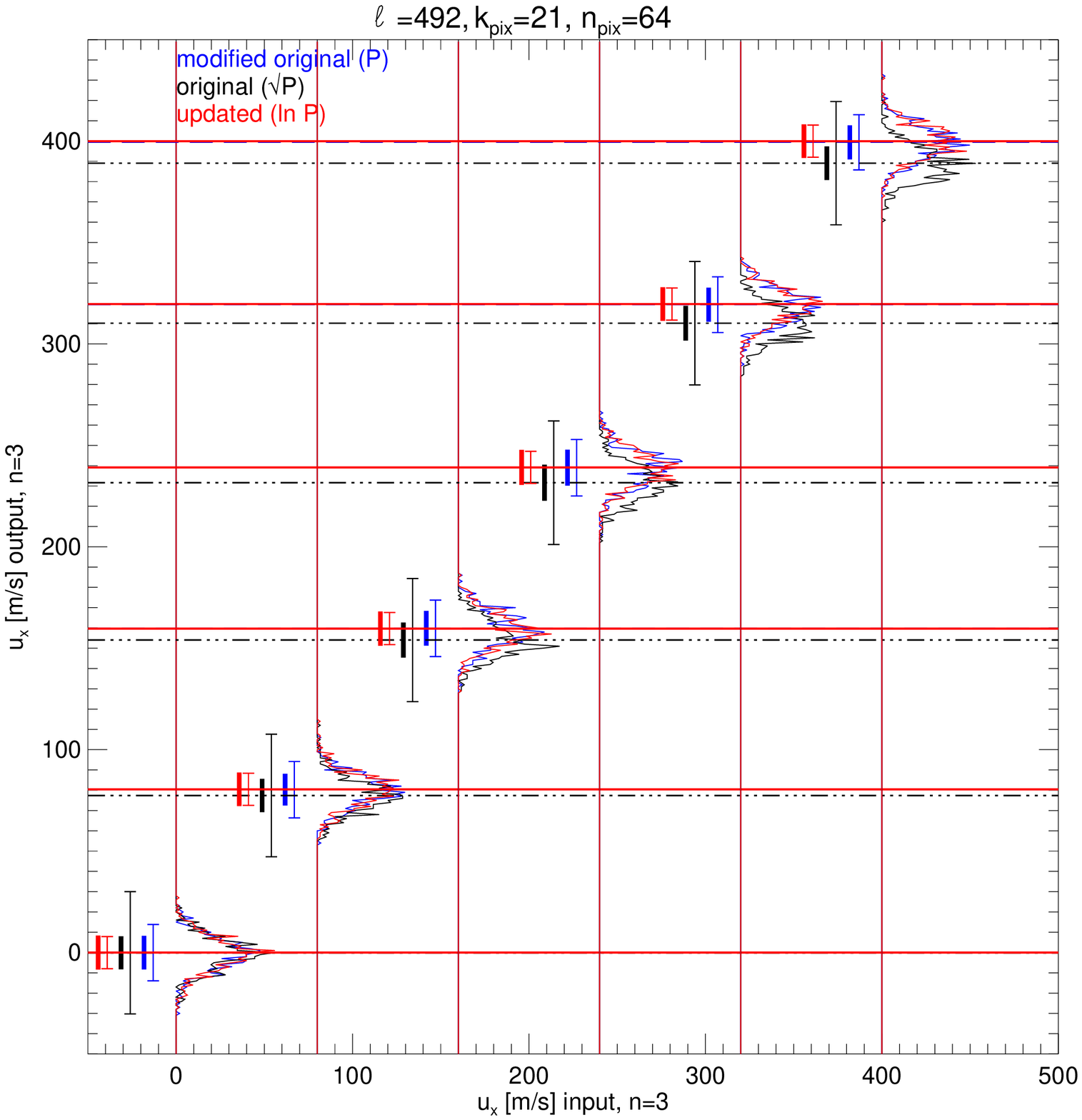}
\caption{
 Fitting results of $u_{x, n=3}$ at $\ell=492$ $k_{\mathrm{pix}}=21$) in the form of normalized histograms by the original code which fit the model to the square root of power (black), the modified original code which fit the power (blue), and the updated code (red). Horizontal lines indicate the average of 500 Monte-Carlo realizations given by the three codes (in the same colors as the histograms, dash-dotted black, dashed blue, and solid red). The short vertical lines on the horizontal lines indicates the error estimated by the codes (thin lines with short horizontal bars) and the standard deviations of the 500 fitting results (thick lines) in the same color as the histograms centered at the means (horizontal lines) by the three codes. We note that the fitting results with overly large deviation are omitted. See text for details.  
 }
\label{fig:ux_iso_inoutcomp_hist2_n3}
\end{figure*}

Figure~\ref{fig:ux_iso_inoutcomp_hist2_n3} shows the fitting results for $u_{x,n=3}$ 
at $\ell=492$ ($k_\mathrm{pix} = 21$)
in the form of normalized histograms of 500 Monte-Carlo simulations. The input models have $q_{3,n}=u_{x,n}=0, 80, \dots  400$~\mps for the $n=3$ ridge. 
Before plotting, we removed outliers from the output fit parameters.  
We removed outliers as described in Sect.~\ref{subsec:error_estimate}.
After the outlier removal procedure, 
the sample numbers for each methods are 394-466 depending on the value of $u_{x,n=3}$ (79-93\%, modified original), 
436-488 (87-98\%, original), and 499-500 (100\%, updated). 
The updated code provides almost no outliers, while
there are some outliers in the fitting results by the modified original and original codes.
The number of outliers depends on $u_{x,n}$, although there is no clear trend with $u_{x,n}$; for example, we cannot say that the stronger flow produces more outliers.
We did not further investigate the details of the outliers in the fitting results by the original and modified original codes.

Figure~\ref{fig:ux_iso_inoutcomp_hist2_n3} shows that
the original fitting code underestimates the input flow by about 3\%. This trend is consistent with what was reported by \cite{2014SoPh..289.2823G}.
The modified original and updated codes produce less-biased flow estimates.
Figure~\ref{fig:ux_iso_inoutcomp_hist2_n3} also compares the errors estimated by the code and the scatter in the Monte Carlo simulation.  The errors from the updated code are consistent with the scatter of the Monte-Carlo results, while the original code overestimates the errors.

\subsubsection{Correlations between the fitting  parameters} \label{subsec:res_cor}

\begin{figure*}[t]
\sidecaption
\includegraphics[width=0.48\textwidth]{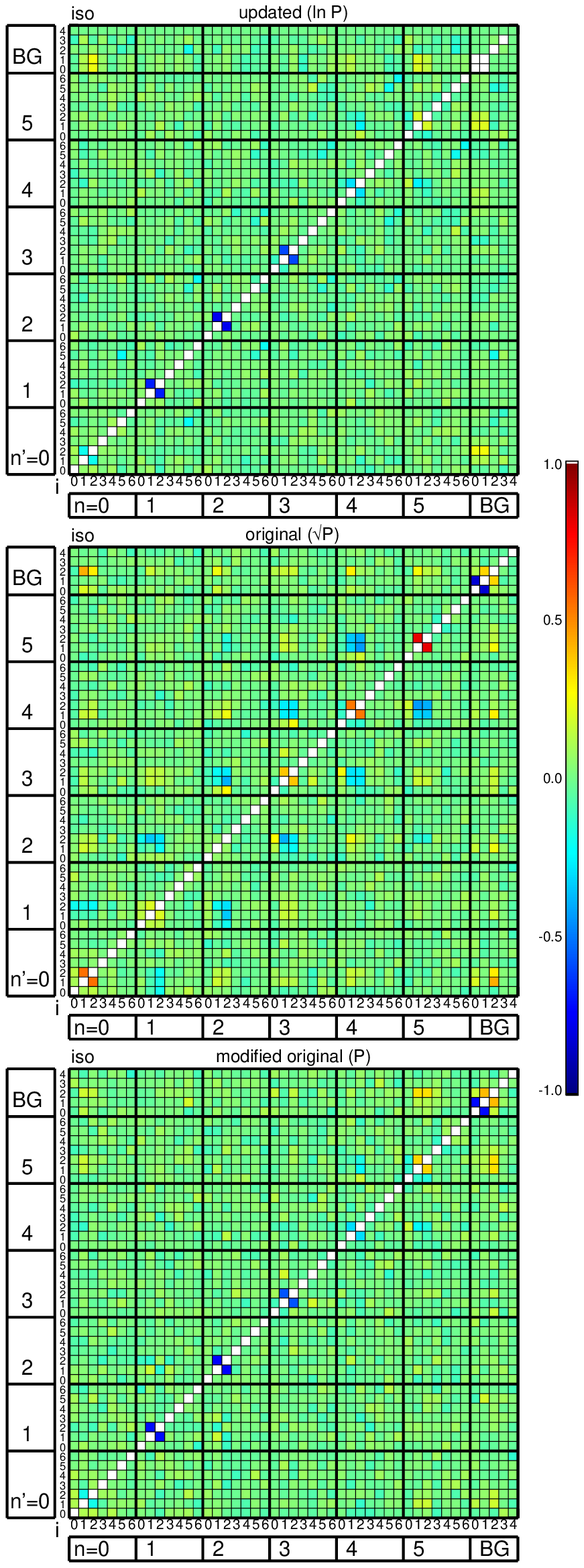}
\caption{
   Correlation coefficients between the parameters (500 Monte-Carlo realizations) for the isotropic model ($u_{x, n} =0$ for all $n$) and for the isotropic model except with $u_x=400$~\mps for the $n=3$ peak (on the next page) at $\ell=492$ ($k_{\mathrm{pix}}=21$). Each box indicates the parameters for $n$-th peak ($n=0,1, \dots 5$), $q_{i,n}$ ($i=0, \dots  6$, from left to right and from bottom to top in each box) and the background parameters (BG). As we defined in Sect.~\ref{eq:altered_Pmodel} $q_{0,n}=\nu_n$ is the frequency of the $n$-th peak, $q_{1,n}=A_n$  is the amplitude, $q_{2,n}=\Gamma_n$ is the width, 
$(q_{3,n},q_{4,n})=\vec{u} = (u_{x,n},u_{y,n})$ is the horizontal velocity, $q_{5,n}=f_{c,n}$ and $q_{6,n}=f_{s,n}$ are parameters to handle anisotropy. Background parameters are 
$q_{0,\mathrm{BG}} = B_0$ is the amplitude,  $q_{1,\mathrm{BG}}=b$ is the power-law index,
$q_{2,\mathrm{BG}}=f_{c, \mathrm{bg}}$ and $q_{3,\mathrm{BG}}=f_{s,\mathrm{bg}}$ are the parameters to handle the anisotropy of the background. Color scale is shown by the color bar on the right side. 
    We note that the last columns and rows ($i=4$) of BG on the top panels are empty because the number of the background parameters in the model in the updated code is four, instead of five.
    }    \label{fig:cor}
\end{figure*}

Figure~\ref{fig:cor} shows the correlation coefficients between the fitting parameters of 500 Monte Carlo samples. For this computation, outliers were removed as described in Sect.~\ref{subsec:error_estimate}.
The original and modified original codes both produce stronger correlations in some of the output parameters
than the updated code.  In particular, for the original code, these parameters show the strongest correlations:
the amplitude $A_n$ ($\qp_{1,n}$) and the width $\Gamma_{n^{\prime}}$ ($\qp_{2,n^{\prime}}$) at peaks $|n-n^{\prime}|\le 1$, 
the roll-off frequency $\nu_{bg}$ ($\qp_{1,\textrm{BG}}$) and the index $b$ ($\qp_{2,\textrm{BG}}$) of the background,  and 
the  index $b$ ($\qp_{2,\textrm{BG}}$) of the background and $A_n$ ($\qp_{1,n}$) or $\Gamma_{n}$ ($\qp_{2,n}$) of higher-$n$ peaks.

The modified original code and the updated code did not show such strong biases, except for the width and amplitude of each peak, along with the background index and some weaker peaks (smaller and larger $n$).
In terms of correlations between the parameters, the updated code shows an overall improvement in comparison with the original,
but the correlation coefficients between $u_x$ and $u_y$ on the same peaks or on the peaks next to each other are $\sim 0.1$ at most for any $u_x$ and any fitting results by all three codes shown in Fig.~\ref{fig:cor}.

\begin{figure}[t]
\includegraphics[width=0.48\textwidth]{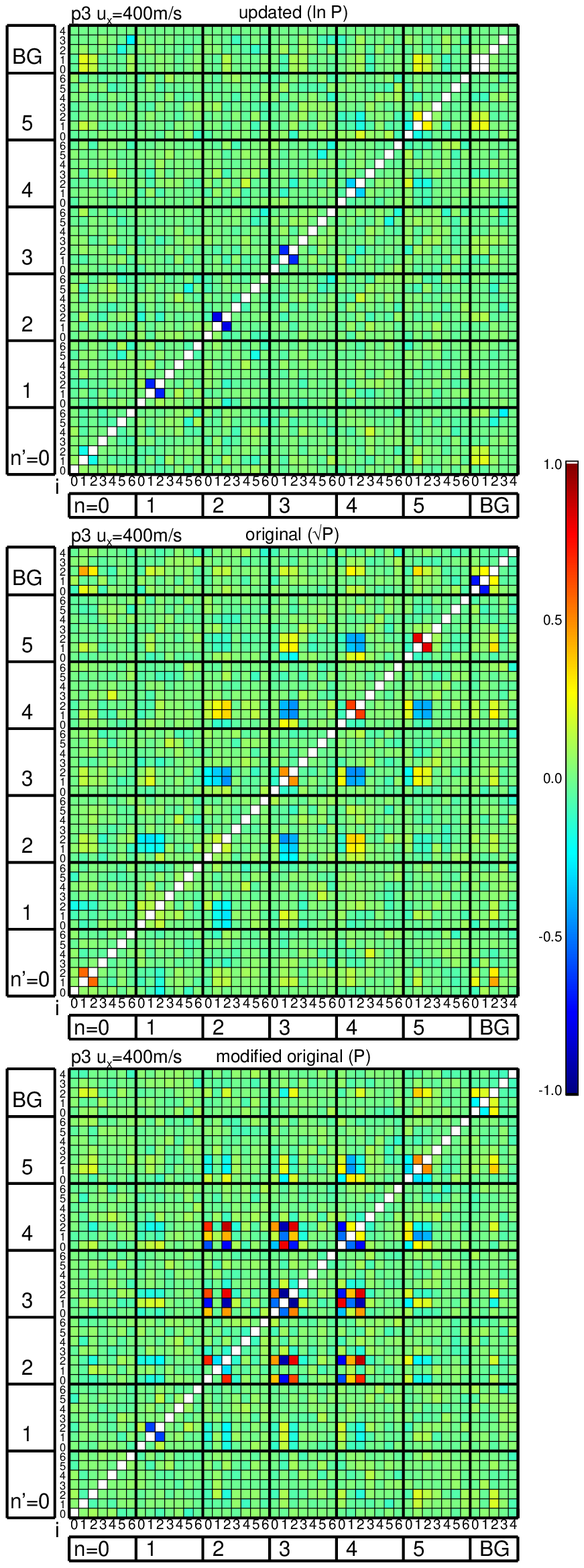}
Fig.4. continued
\end{figure}

\subsection{Summary of the Monte-Carlo test}

The Monte-Carlo tests show
that the updated code is able to reasonably recover
the parameters, apart from the amplitude, that are used to generate the input power spectra.
Other than the amplitudes, we were not able to measure
a statistically meaningful bias in the fit results using 500 Monte-Carlo simulations.
The underestimation of the amplitude is the result of taking the logarithm of the rebinned power.  Appendix~\ref{apsec:app_amp} discusses this issue in detail. Since current ring-diagram flow inversions are carried out using the mode shifts, $(q_{3,n}, q_{4,n}) = (u_{x,n}, u_{y,n})$, and the other fitting parameters including mode amplitudes are not used, we believe underestimated amplitudes will not substantially impact on further analysis. 

\section{Discussion} \label{sec:discuss}

\subsection{Further rebinning in azimuth} \label{sec:dis_rebin}

We explored the idea of rebinning the data further in $\theta$.
If we rebin further, the PDF of the resulting power spectrum is closer to Gaussian.
Additional rebinning has the additional benefit of reducing the underestimation of the amplitude.
Another benefit is that if we can reduce the number of pixels involved in the fit without a significant decrease of the fitting quality, it 
will reduce calculation costs.

\begin{figure*}[hp]
\centering
\includegraphics[width=0.75\textwidth]{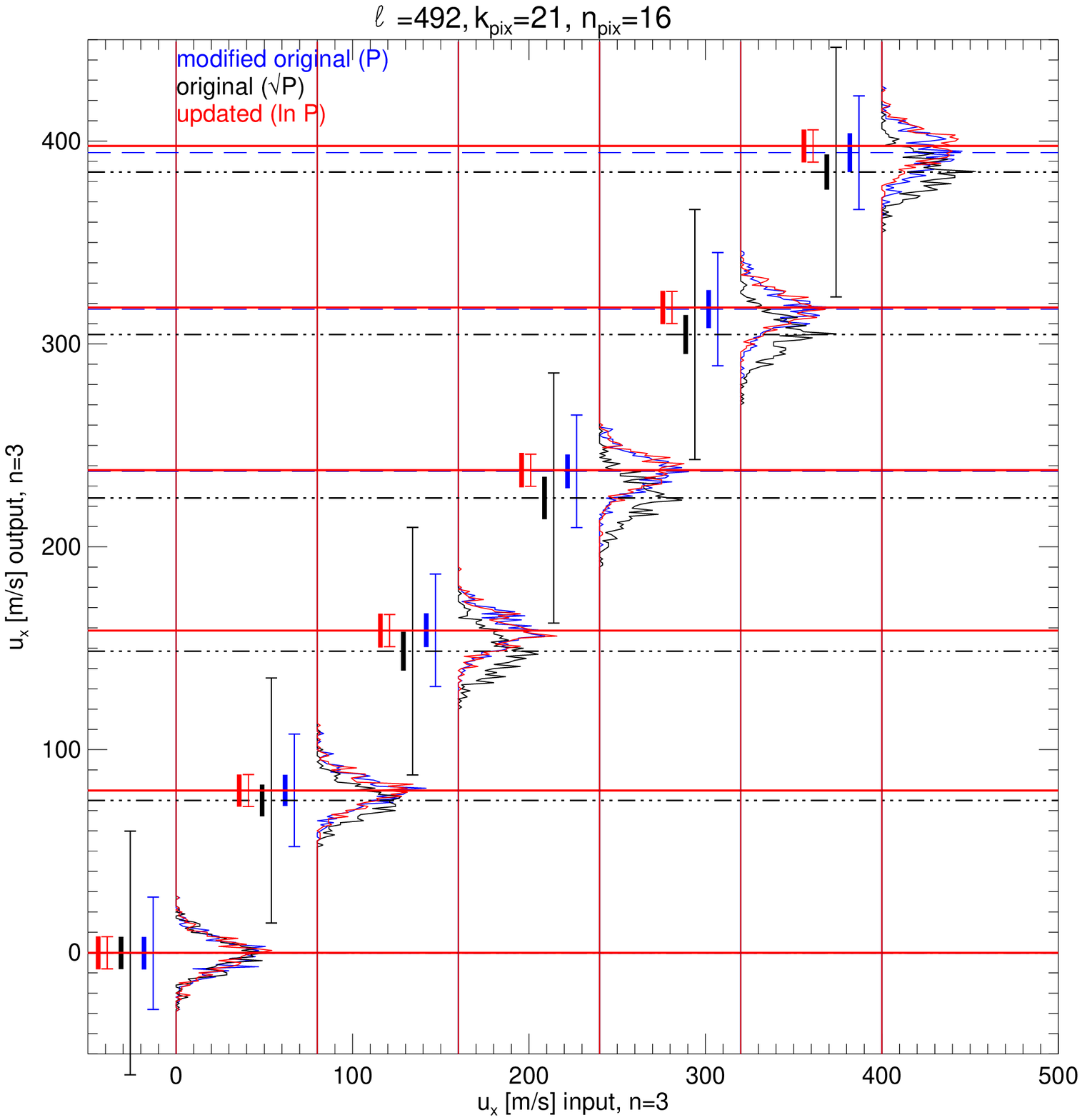}
\caption{Fitting results of $u_{x,n=3}$ at $\ell=492$ ($k_{\mathrm{pix}}=21$), similar to Fig.~\ref{fig:ux_iso_inoutcomp_hist2_n3}, but for the further rebinned data. In this case, the number of azimuthal pixels, $n_{\mathrm{pix}}$ is reduced to 16, instead of the original 64. Legends and colors are similar to those of Fig.~\ref{fig:ux_iso_inoutcomp_hist2_n3}. Scale factors for the error estimates by the updated code are adjusted accordingly.}
\label{fig:ux_iso_inoutcomp_hist2_n3_reb}
\end{figure*}

Figure~\ref{fig:ux_iso_inoutcomp_hist2_n3_reb} shows the fitting results corresponding to
the four-time further rebinning. Specifically, we rebin the 64-pixel azimuth grid into 16~pixels.  
In this case, valid sample numbers for the three codes are 351-429 (70-86\%, modified original), 406-487 (81-97\%, original), and 500 (100\%, updated).
While the updated code had no outliers, the numbers of outliers in the results by the original and modified original codes
increase compared to the case without further rebinning (see Sect.~\ref{subsec:flow_singlepeak}).
This also confirms the stability of the updated code. 

This figure tells us that our updated code is better than the original for this further rebinned case as well, in terms of a more reasonable error estimate and a smaller bias. 
Moreover, using our updated code, we can rebin further up to 16 pixels at this $k$ ($k_\mathrm{pix}=21$)  without a significant increase in the noise or the underestimate of the parameters or outliers in comparison with the original 64-pixel case. 
In the further rebinned case, the correlations between parameters show a trend that is essentially similar to that of Fig.~\ref{fig:cor}. The only exception is in the fitting results obtained by the modified original code; they show less correlation even in the case of $u_{x,n=3}=400$ \mps in the further rebinned case.

In the original code, there is no scaling factor related to  the bin number and the error estimate by the code is needed to be scaled properly.  
For these 64-pixel and 16-pixel cases, the scatters are not significantly changed and it is the case as well for the error estimated with the proper scaling.
For these plots, again we note that we need a sufficient number of pixels for a good fitting; in this case 8 or 4 bins were too small, because the functional form of the model has terms that vary as $\cos \theta$ and $\sin \theta$ but also as $\cos 2\theta$ and $\sin 2 \theta$ as parameters.

\subsection{Future work} \label{subsec:disc_tasks}

The modifications described in this work are mainly corrections of problems in the original code.
The exception to this is the change of the algorithm from the maximum likelihood method based on the chi-square distribution function to the one based on the normal distribution function, namely the least-squares method, and the fitting not to the square-root of power but to the logarithm of the power. 
There are several potential improvements of the analysis.
While implementing such further alterations are beyond the scope of this paper,
we will briefly discuss some potential future improvements.

One of the open issues is the remapping and rebinning.
Currently the power is remapped from Cartesian to polar coordinates and then smoothed in the azimuthal direction. 
However, it is possible to do the analysis in the original Cartesian system 
as it is done in the HMI pipeline {\bf fitsc} module \citep{2011JPhCS.271a2008B}, which fits
in slices at constant temporal frequency.  
The fitting approach shown here would only
need to be slightly modified to be carried out in a region with $\sqrt{k_x^2 + k_y^2}$ near $k$.
We expect that the main issue would be extending the model to allow for small variations in $k$.
As most parameters are presumably smooth in $k$, we speculate the fitting small range in the $k$ rather
than single $k$ would help the stability of the fits.

In the current code, we approximate the PDF of the remapped and smoothed power spectrum as a normal distribution function.
But how we rebin, including the topic mentioned in Sect.~\ref{sec:dis_rebin}, is still open. 
In the method shown here, we do not rebin the model function 
but use the model function calculated on the  same grid as the rebinned data. 
This is an approximation in the limit where the rebinning does not significantly change
the shape of the limit spectrum and we expect that it will cause a bias in the 
case of extreme rebinning.
Figures~\ref{fig:ux_iso_inoutcomp_hist2_n3} and~\ref{fig:ux_iso_inoutcomp_hist2_n3_reb} show
that the bias in the fitting results of the updated code is less than about $5$~\mps
in the range of $|u_x| \le 400$~\mps even in the further rebinned case (Fig.~\ref{fig:ux_iso_inoutcomp_hist2_n3_reb}).

The choice of the model function (currently Eq.~(\ref{eq:altered_Pmodel})) is also an open issue.
As shown in Fig.~\ref{fig:modelfunc}, the current model function does not reproduce the detailed structure of the observed power spectrum. 
For example, the current model function does not take into account
the asymmetry of the ridge shape in terms of frequency.
Currently, we use the power spectrum of the tile at the disc center only 
but that is the simplest case and when we investigate the deep convection,
we cannot avoid using the tiles on various locations on the disc. 
In such a case, in order to construct a model function, 
we need to take into account further effects, such as the effect of the center-to-limb variations \citep{2012ApJ...749L...5Z}, the line-of-sight effect on shape of the power,
or the effect of the Postel projection. 
Also, the effects of differential rotation on eigenfunctions are to be accounted for in future.

\section{Conclusions} \label{sec:conc}

We identified several problems  in the multi-ridge fitting code \ATLAS \citep{2014SoPh..289.2823G}
and we updated the code in response. We  confirmed that flow-estimate biases and error overestimates exist in the fitting results by the original code.  The biases that we found are insufficient to 
resolve the discrepancy presented in \cite{HGS2016}.

The updated code is based on a consistent model and an appropriate likelihood function.
Monte Carlo tests show the fitting results and their error estimates by the updated code are reasonable
and confirm the improvement of the fitting.

The work shown here is limited to the fitting part of the ring-diagram analysis codes. 
The next step in the ring-diagram analysis is flow inversion using the mode-shift parameters ($q_{3,n}$ and $q_{4,n}$ in Eq.~(\ref{eq:altered_Pmodel})). Unlike the HMI ring-diagram pipeline in which the inversion is done at each tile, \cite{2015ApJ...803L..17G} used a 3-D inversion using multiple tiles. This unique step should be be the focus of examination in future works. 

\begin{acknowledgements}
We thank Rick Bogart for useful discussions.
The German Data Center for SDO (GDC-SDO), 
funded by the German Aerospace Center (DLR),
provided the IT infrastructure to process the data.
The HMI data used are courtesy of NASA/SDO and the HMI science teams.  
BWH acknowledges NASA support through grants 80NSSC17K0008, 80NSSC18K1125, and 80NSSC19K0267.
We acknowledge partial support from ERC Synergy grant WHOLE SUN 810218.
\end{acknowledgements}




\appendix 

\section{Original \ATLAS code}\label{apsec:original_code}

\subsection{Fitting method of the original \ATLAS code} \label{apsec:original_method}
In the original \ATLAS code \citep{2014SoPh..289.2823G, 2015ApJ...803L..17G}, 
the processing steps after computing the power spectrum for each tracked tile are as follows:
\begin{enumerate}
\item at each frequency $\nu$, remap the log of the square root of the power spectrum $\ln\sqrt{P(\bk,\nu)}$ from the Cartesian $(k_x,k_y)$  to polar ($k,\theta$) coordinates, 
\item compute  $\exp{\ln\sqrt{P(k,\theta,\nu)}}$ and carry out Fourier interpolation using a box-car low-pass filter  to smooth in the azimuthal ($\theta$) direction. The number of the grid points in azimuth $\theta$ is reduced  from $n_{\mathrm{pix}}=256$ to $n_{\mathrm{pix}}= 64$, 
\item fit a Lorentzian model (see Appendix~\ref{apsec:original_model}) to the smoothed square-root power at each $k$.
The fitting is done  based on the maximum likelihood method assuming that the PDF of the power is the chi-square distribution with two degrees of freedom (see more details in Appendix~\ref{apsec:pdf_raw}), and
\item estimate error of the fitting parameters separately from the fitting using the Fisher information matrix based on the final fitting results.
\end{enumerate}

\subsection{Model for the power spectrum in the original \ATLAS code} \label{apsec:original_model}

In the original \ATLAS code \citep{2014SoPh..289.2823G, 2015ApJ...803L..17G}, the power at each $k$ is modeled with the sum of Lorentzians with seven parameters ($q_{i,n}^\prime, i=0,\dots 6$) for each ridge $n=0, \dots  \nr-1$, where $\nr$ is the number of ridges, 
and a background model with five parameters ($q_{i,\mathrm{BG}}^\prime, i=0,\dots 4$) at each $k$.  The total number of fit parameters at each $k$ is $7 \nr + 5$.

The model power spectrum, at a single $k$, is:
\begin{eqnarray}
&&P^{\prime}(\theta,\nu; \bq^\prime_\mathrm{BG}) = \nonumber \\
&& \sum^{n_{\mathrm{r}}-1}_{n=0} \frac{\qp_{1,n} (\qp_{2,n}/2) F^\prime(\theta ;\qp_{5,n}, \qp_{6,n})}{[\nu - \qp_{0,n} +
k (\qp_{3,n} \cos \theta + \qp_{4,n} \sin \theta) /(2\pi)]^2+(\qp_{2,n}/2)^2}  \nonumber \\
&&+ \ B^\prime(\theta,\nu; \bq^\prime_\mathrm{BG}) , \label{eq:Pmodel}
\end{eqnarray}
where 
\begin{equation}
F^\prime(\theta; p_0, p_1 )= 1 +p_0\cos[2(\theta-p_1)] , \label{eq:Pmodel_anis}
\end{equation}
and $\qp_{0,n}=\nu_n$ is the frequency of the $n$-th peak, $\qp_{1,n}=A_n$  is the amplitude, $\qp_{2,n}=\Gamma_n$ is the width, 
$(\qp_{3,n},\qp_{4,n})=\vec{u} = (u_{x,n},u_{y,n})$ is the horizontal velocity, $\qp_{5,n}=f_n$ and $\qp_{6,n}=\theta_n$ are anisotropy terms in Eqs.(5) and (6) in \cite{2014SoPh..289.2823G}.
The background is modeled at each $k$ as 
\begin{eqnarray}
B^\prime (\theta,\nu; \bq^\prime_\mathrm{BG}) = \frac{\qp_{0,\mathrm{BG}}}{1+(\nu/ \qp_{1,\mathrm{BG}})^{\qp_{2,\mathrm{BG}}}} F^\prime(\theta ; \qp_{3,\mathrm{BG}}, \qp_{4,\mathrm{BG}})  , \label{eq:Bmodel}
\end{eqnarray}
where 
$\qp_{0,\mathrm{BG}} = B_0$ is an amplitude, 
$\qp_{1,\mathrm{BG}}=\nu_{\mathrm{bg}}$ is a roll-off frequency, 
$\qp_{2,\mathrm{BG}}=b$ is the power-law index,
$\qp_{3,\mathrm{BG}}=f_{\mathrm{bg}}$ is the amplitude of the anisotropy term, and 
$\qp_{4,\mathrm{BG}}=\theta_{\mathrm{bg}}$ is the phase of the anisotropy term 
in Eq.~(7) in \cite{2014SoPh..289.2823G}.

\subsection{Problems in the original \ATLAS code}\label{apsec:issues}
First, the Lorentzian model for the power in the original code was fit
to the square root of the observed power. This is inconsistent with what is stated in \cite{2014SoPh..289.2823G,2015ApJ...803L..17G}
and it is also an inconsistency between the model and the observable. 
Therefore, we made them consistent in the updated code (see Sect.~\ref{subsec:LSfitting}).

Second,  the cost function to be minimized in the original code was not a good approximation. 
The cost function to be minimized in the maximum likelihood method based on the 
chi-square-distribution with two degrees of freedom as PDF is 
\begin{equation}
    - \ln \mathcal{L} (\bq) = \sum_i \left\{ \ln\left(\frac{P_i (\bq)}{O_i}\right)+ \left(\frac{O_i}{P_i  (\bq)}\right) \right\} \ ,
\end{equation}
where  $O_i$ is the observed spectrum and $P_i$ is the model, 
as given in Sect.~\ref{apsec:original_model}. The sum is taken over all grid points 
$(\theta_i, \nu_i)$ in the fitting range.

Instead, 
the original code minimizes
\begin{equation}
\sum_i  \mathcal{B}_i^2 \equiv \sum_{i} \left\{ \ln\left(\frac{P_i (\bq)}{O_i}\right)+ \left(\frac{O_i}{P_i (\bq)}\right)  \right\}^2 
\label{eq:MRF_dev}
\end{equation}
using the mpfit.c code\footnote{This library code minimizes the sum of square of given function, and $\mathcal{B}_i$ in Eq.~(\ref{eq:MRF_dev}) is given in the original \ATLAS code.}. 
This is conceptually 
inconsistent with the description in \cite{2015PhDT.......167G}, although the likelihood function itself is not explicitly stated there.  
While it can be shown that the results of minimizing of the exact cost function (Eq.~(\ref{eq:-lnL}))
and the wrong one (Eq.~(\ref{eq:MRF_dev})) are identical in the limit of linear perturbations,
there is no reasonable computational or physical reason to take the extra square in the calculations. Moreover, it is not useful for the error estimates. Therefore, we decided to correct this issue; Sect.~\ref{subsec:LSfitting} describes our changes. 

Third, 
in the original \ATLAS code, error estimates are obtained by an independent calculation of 
the Fischer information using the final fitting results and using the chi-square distribution with two degrees of freedom as the likelihood function.
It is not consistent to compute the error estimates with a likelihood function that is different than what was used in the fitting itself.  In Sect.~\ref{subsec:LSfitting}, we also describe a consistent approach to the computation of error estimates in our updated code. 

Fourth, several parameters are unstable. Therefore, we have changed the parameterizations as we discussed in Sects.~\ref{sec:alt_anis} and \ref{sec:alt_backparams}. 

\section{Probability distribution function (PDF) and maximum likelihood method based on the PDF} \label{apsec:PDF}
\subsection{PDF of the raw power spectrum} \label{apsec:pdf_raw}

\cite{1984PhDT........34W} demonstrated that the PDF of
a single observed power spectrum divided by 
the expectation value of the spectrum is the chi-square distribution 
with two degrees of freedom. 
On this basis,
\cite{1986ASIC..169..105D} and \cite{1990ApJ...364..699A} introduced the 
probability density at a given grid  point in the Fourier space $(k_x, k_y,\nu)_i$  
\begin{equation}
\mathcal{P}(O_i) = \frac{1}{P_i (\bq)} \exp \left(-\frac{O_i}{P_i (\bq) }\right) ,
\end{equation}
where $O_i$ is the observed spectrum, 
$P_i$ is the model for the limit spectrum, and $\bq$ is the model parameters. 
The joint probability density for the experimental outcome at horizontal independent wavevectors and  frequencies is given by 
\begin{equation}
\mathcal{L}=\prod_i  \mathcal{P}(O_i)= \exp\left[ -\sum_i \left(\ln P_i (\bq) +\frac{O_i}{P_i(\bq) } \right) \right] \  .
\end{equation}
This is the likelihood function, and the model parameters which maximize this $\mathcal{L}$ are the targets in the maximum likelihood method. 
In practice $- \ln \mathcal{L}$ is minimized to use standard minimization procedures.
To make the computation simpler, $\ln O_i$ is subtracted from $-\ln \mathcal{L}$.
The minimization of $-\ln \mathcal{L} - \ln O_i$ in terms of \bq \ is identical to that of $-\ln \mathcal{L}$,
because $O_i$ is not a function of the model parameter \bq. 
In conclusion, 
\begin{equation}
-\ln \mathcal{L} - \ln O_i = \sum_i \left(\ln \frac{P_i  (\bq) }{O_i } +\frac{O_i}{P_i (\bq)} \right) 
\label{eq:-lnL} 
\end{equation}
is minimized.

\subsection{PDF of logarithm of averaged power} \label{apsec:pdf_logP}

The  PDF of a well-averaged power spectrum is also a chi-square distribution but with many more than the two degrees of freedom of the original and given the central limit theorem, it can be approximated by a normal distribution.
To carry out the least-squares fitting based on the normal distribution function and estimate errors of the fitting results, we need the variance of the normal distribution function. 
We therefore take advantage of the fact 
that the logarithm of the averaged spectrum obeys a normal distribution function with a constant variance. 
In Appendix~\ref{apsec:pdf_logP_derive}, we show that if we have a spectrum obtained by averaging over $N$ spectra, each normally distributed with the same expectation value $M$,
 the logarithm of averaged spectra, $y$, obeys the normal distribution function $N(\ln M, \sigma_N^2)$:  
\begin{equation}
f(y) = \frac{1}{\sqrt{2\pi }\sigma_N} \exp{\left[-\frac{1}{2} \left(\frac{y-\ln M}{\sigma_N}\right)^2\right]}\ , \label{eq:logP_pdf}
\end{equation}
where $\sigma_N = 1/ \sqrt{N}$,  therefore, $\sigma_N$ is a constant over $y$. 
We note that the function $f(y)$ is linearized around the mean to derive Eq.~(\ref{eq:logP_pdf}); see Appendix~\ref{apsec:pdf_logP_derive} for details.

In the present case, we fit one $k$ at a time. To do this, the spectra are, at each frequency, linearly interpolated in $k_x$ and $k_y$ to a circle with radius $k_{\rm pix}$, where $k_{\rm pix}$ is $k$ in units of bins, and smoothed to $n_{\rm pix}$ azimuths. On this basis, the question is if we can still use a least-squares fit with a diagonal covariance and a single $\sigma_N$ and, if so, which value of $\sigma_N$ (or equivalently $N$) in Eq.~(\ref{eq:logP_pdf}) should be used. In particular, it needs to be considered that more than $2\pi k_{\rm pix}$ are used for the interpolation and that the averaged values will be correlated.

\begin{figure}[]
\centering
\includegraphics[width=0.48\textwidth]{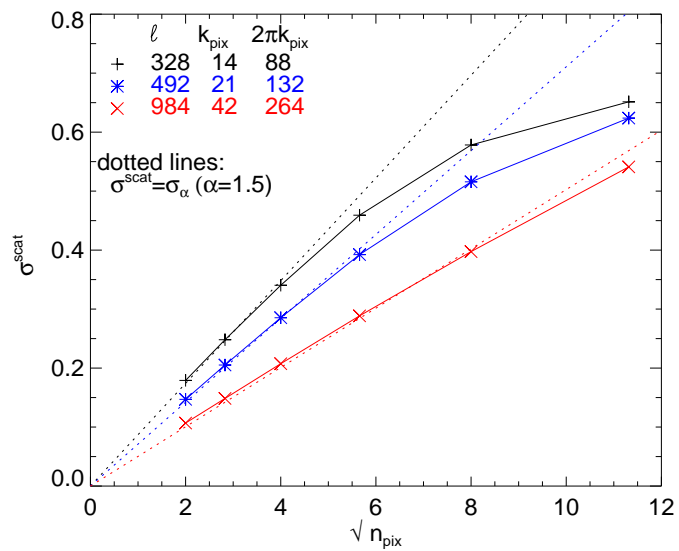}
\caption{Standard deviation of the logarithm of chi-square distributed white noise rebinned on the annulus with the radius of $k_{\mathrm{pix}}$ pixels, 
 $\sigma^{\mathrm{scat}}$, is plotted against the square-root of the pixel number on the annulus after rebinning, $\sqrt{n_\mathrm{pix}}$, at three $k_\mathrm{pix}$ shown with different symbols in different colors.   
 In the case of $n_\mathrm{pix} \ll 2 \pi k_\mathrm{pix}$, $\sigma^{\mathrm{scat}}=\sigma_\alpha ( \alpha=1.5$), which is shown with the dotted lines, is a valid approximation, while in the limit of $n_\mathrm{pix} \gtrsim 2 \pi k_\mathrm{pix}$,
 $\sigma^{\mathrm{scat}}$ deviates from $\sigma_\alpha$ and approaches a constant of $2/3$. 
}
\label{fig:whitenoise_sigma}
\end{figure}

In the limit of averaging over the entire circle, namely $n_{\rm pix}=1$, and in the case with $k_{\rm pix} \gg 1$, it might be reasonable to assume that $N=2\pi k_{\rm pix} \alpha$, with $\alpha$ accounting for the fact that the averaging is essentially over an annulus around $k_{\rm pix}$. Indeed, a Monte-Carlo test shows that $\alpha \approx 1.5$ gives a very good estimate of the error on the average, which will be shown in Fig.~\ref{fig:whitenoise_sigma} later in this subsection.

For $n_{\rm pix}\gg 1$ the variances, as well as the off-diagonal elements of the covariance matrix of the interpolated datapoints  will, in general, depend on azimuth.
Assuming that the fitted function may be linearized in the fitted parameters, a linear fit implies that the fitted parameters are given by a linear combination of the observed values.  Assuming that the functions are smooth (as in the present case where they are low-order harmonic functions), the coefficients in the linear combination are also smooth. From this it follows that if the variations in the properties of the covariance matrix average on the scale of the variations in the fitted functions, then the same scaling factor may be used and that
\begin{equation}
   N \sim \frac{ 2 \pi k_{\mathrm{pix}} \alpha }{n_\mathrm{pix}} \ ,  \label{eq:N_alpha}
\end{equation}
which a Monte-Carlo test again confirms.

To validate the arguments above, 
a simple Monte-Carlo test 
 is carried out to measure the scatter (standard deviation, $\sigma^{\mathrm{scat}}$ ) of the logarithm of averaged white noise.  
 White-noise fields with the chi-square distribution with two degree of freedom in a  three-dimensional Cartesian Fourier space $(k_x,k_y, \nu) = (384, 384, 1152) \mathrm{[pix]}$ are remapped at several annuli with specific radii $k_\mathrm{pix}$ in the same way as the data in the updated analysis code. 
 At first, there are 256 pixels on each annulus after remapping, we then rebin them into $n_\mathrm{pix} = 4,8, 16, 32, 64, \mathrm{and} \ 128 $ pixels. The standard deviation of the logarithm of the rebinned white noise,
 $\sigma^{\mathrm{scat}}$, are computed and plotted against the square root of the  number of datapoints after rebin on the annuli, $n_\mathrm{pix} = 14, 21, \mathrm{and} \ 42$
 in Fig.~\ref{fig:whitenoise_sigma}.
As we mentioned above, in the case of $n_\mathrm{pix} \ll 2 \pi k_\mathrm{pix}$, $\sigma^{\mathrm{scat}} \simeq \sigma_\alpha (\alpha=1.5) $, where 
$\sigma_\alpha ( \alpha) \equiv \sqrt{n_{\mathrm{pix}}/(2 \pi k_{\mathrm{pix}} \alpha)} $. 
In the case of $n_\mathrm{pix} \gtrsim 2 \pi k_\mathrm{pix}$ the deviation of $\sigma^{\mathrm{scat}}$ from $\sigma_\alpha$ is not negligible,
and $\sigma^{\mathrm{scat}}$ approaches a constant $2/3$; this constant can be derived from the bi-linear interpolations of chi-square distributed random variables in two dimensions.
Also we note that 
we have not included the apodization effect in the data to create the noise field in this test calculation.
Without apodization $\sigma^{\mathrm{scat}}$ might be overestimated;
in any case, this does not affect the parameter estimates but only the error estimates. 

In our updated code, $\sigma_N$ is set to $1$ (see the cost function Eq.~(\ref{eq:-lnL2})), as we mentioned in Sect.~\ref{subsec:LSfitting},
 and after the fitting is done the error provided by the code is scaled by $\sigma_\alpha$.
Although calling the minimization code with an incorrect error estimate can lead to worse convergence, the difference in this case should be negligible. Implementing $\sigma_\alpha$ in the code is a  task for the future.

\subsection{Derivation of PDF of the logarithm of averaged power}\label{apsec:pdf_logP_derive}

In this section we briefly summarize how to derive the PDF of the logarithm of the power. 

Suppose $x$ is the average of $N$ spectra whose averages are $M$. If $N$ is large enough, then the PDF of this spectrum is the normal distribution 
\begin{equation}
f_x(x) = \frac{1}{\sqrt{2\pi}\sigma} \exp {\left[-\frac{1}{2} \left(\frac{x-M}{\sigma}\right)^2\right]} 
\end{equation}
and $\sigma=M/\sqrt{N}$. 
Here we define 
\begin{equation}
y=g(x) \equiv \ln x  
\end{equation}
and when $x$ is close to $M$, 
\begin{equation}
y =\ln\left[M\left(1+\frac{x-M}{M}\right)\right] \simeq \ln M + \frac{x-M}{M} .
\end{equation}
Therefore, the inverse function of $g$ is given by 
\begin{equation}
g^{-1} (y)= x = (y-\ln M) M+M
\end{equation}
and 
\begin{equation}
\frac{\mathrm{d}}{\mathrm{d}y} (g^{-1} (y)) =M .
\end{equation}
Hence, the PDF of $y=\ln x$ is given by 
\begin{eqnarray}
f_y(y) &=& \left|\frac{\mathrm{d}}{\mathrm{d}y} (g^{-1} (y))\right|f_x(g^{-1}(y)) \nonumber \\
&=& \frac{1}{\sqrt{2\pi}\sigma_y } \exp{\left[-\frac{1}{2} \left(\frac{y-\ln M}{\sigma_y}\right)^2\right]} \ ,
\end{eqnarray}
where $\sigma_y = \sigma/M = 1/\sqrt{N}$.
Therefore, the logarithm of the averaged power has a constant error, which is independent of the original $\sigma$ and depends only on the number of the averaged spectra, $N$.
For  comparison, approximated PDF and the distribution functions of the Monte-Carlo samples are given in Appendix~\ref{apsec:dist_func_comp}.

\subsection{Comparison of the distribution functions with Gaussian and chi-square distribution functions} \label{apsec:dist_func_comp}

In this subsection we show that Gaussian function is a good approximation for the distribution functions of the logarithm of the synthesized power.  We also discuss the degrees of freedom of the chi-square distribution functions of the unwrapped and smoothed power here.

\begin{figure*}[hbtp]
\centering
\includegraphics[width=.95\textwidth]{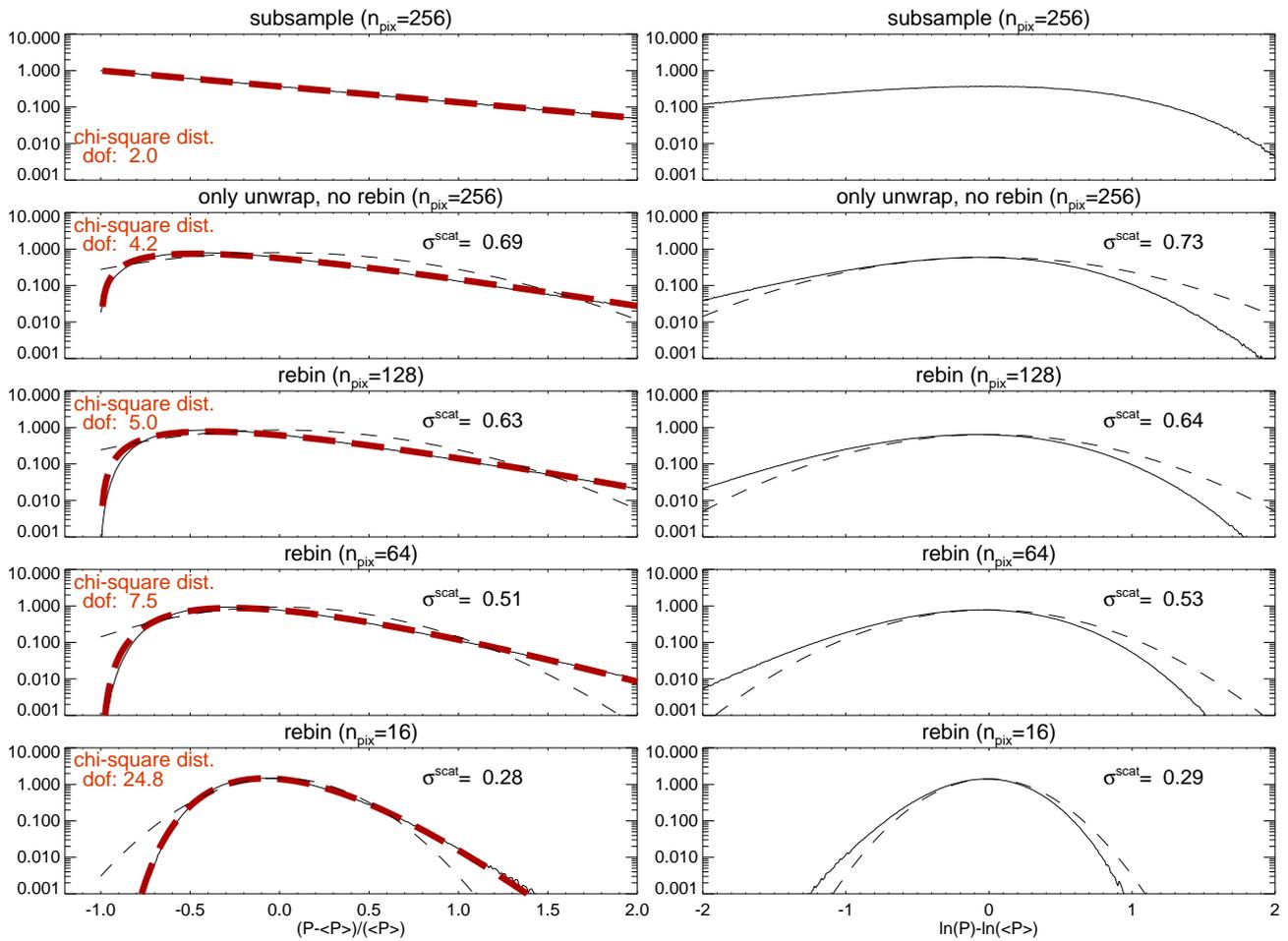}
\caption{
 Normalized frequency distribution functions of the normalized synthetic power at $k_{\mathrm{pix}}=21$ ($\ell=492$, black solid lines): top panels are for the subsampled power and the second panels are the unwrapped power without further smoothing, while the lower panels are for the rebinned powers. Left panels show $(P-\langle P \rangle)/\langle P \rangle$,
and right panels show $\ln P - \ln \langle P \rangle$, where $\langle \rangle$ indicates the expectation value (model). 
 The input model of the power is isotropized ($q_{i,n}=0$ for  $3\le i \le 6$ for all $n$ and $q_{i,b}=0$ for $i=2,3$) but otherwise it is the same as the model with the parameters given in Table \ref{table:input_parameters_k21}.  
 For this plot, we use 50 realizations for the top two rows and 500 realizations for the rebinned data (lower three rows). 
 The black dashed lines on the lower four panels are the Gaussian function centered at zero whose width is the standard deviation of the samples (shown on the panels). This validates our approximation of the distribution function of the logarithm of the power by the Gaussian function.
 The thick red dashed curves on the left panels are the chi-square distribution function of
 various degrees of freedom. The degree of freedom (dof) for each distribution is determined from the standard deviation of 
 each set of normalized power. See the text for details. 
}
\label{fig:log190403_testpdf}
\end{figure*}

Figure \ref{fig:log190403_testpdf} shows the distribution function of the power and the logarithm of the power at $k_{\rm pix}=21$ ($\ell=492$) from the Monte Carlo simulations. 
The Gaussian function centered at zero whose width is the standard deviation of the samples are overplotted in the figure. These plots show that the Gaussian function is a good approximation 
for the distribution function of the logarithm of the 
unwrapped and smoothed power  near the mean value,
though the tails of the distribution function are longer than those of the Gaussian.

In the left panels of Fig.~\ref{fig:log190403_testpdf}, the distribution function of the power is also compared with the chi-square distribution functions with various degrees of freedom: 
the chi-square distribution functions shown here are 
\begin{equation}
    f(X;d)=\left(\frac{d}{2}\right)^{d/2} \frac{(X+1)^{d/2-1}}{\Gamma (d/2)} \exp{\left(-\frac{d}{2} (X+1)\right)} \ ,
\end{equation}
where $X=(P-\langle P \rangle)/\langle P \rangle$, $\langle \rangle$ indicates the expectation value, and $d$ is the number of degrees of freedom (dof).
Since in this case $\langle X \rangle =0$ and the variance of $X$ is $2/d$,   
we determine dof from the variance and show the corresponding chi-square distribution function in the figure. 
As the original synthesized power is created from the model and the two-dof chi-square distributed random numbers, subsampled power is purely two dof. The unwrapped power has a four-dof chi-square distribution; this indicates that 
unwrapped power is effectively the subsampled power rebinned over two pixels. 
Because in this case the radius of annulus is $k_{\rm pix}=21$ and hence roughly $2 \pi k_{\rm pix}$ independent pixels are on the annulus,
the unwrapped power with 256 pixels on the annulus is roughly twice oversampled. 
Therefore, the dof of the power unwrapped and rebinned over two pixels 
in azimuth ($n_{\rm pix}=128$, third panel) is five. This is 
almost unchanged from the original  (second panel, dof=4).  
This is also the case when we further rebin the data over four pixels (fourth panel); dof is only eight and this is smaller than 16, which we might have expected for the four-pixel rebinned four-dof distributed independent variables (i.e., unwrapped power).
But we should also note here that 12 is the dof which we expect on the basis of the effective rebinning numbers we discussed in Appendix~\ref{apsec:pdf_logP} (Eq.~(\ref{eq:N_alpha})), $N \sim 3 $, where $\alpha=1.5$.
It is also larger than the dof of the distribution function (8); 
this is partly because the approximation (Eq.~(\ref{eq:N_alpha})) is not sufficiently good for $n_{\rm pix}=64$ case as was shown in Fig.~\ref{fig:whitenoise_sigma}.
When we see the further rebinned case with $n_{\rm pix}=16$, namely the unwrapped power averaged over 16 pixels 
(the lower-most panel), the dof is 25 and close to what we expect from Eq.~(\ref{eq:N_alpha}), $N\sim 6$: in this case the dof is $6 \times 4  = 24$. For this case  the approximation by Eq.~(\ref{eq:N_alpha}) with $\alpha=1.5$ is good according to Fig.~\ref{fig:whitenoise_sigma}. 
This closeness of the distribution function of the remapped and smoothed power 
to the chi-square distribution with some degrees of freedom suggest that
one might expect improvement to the fitting method  
by using the maximum likelihood method based on the
chi-square distribution with  specific degrees of freedom.

\section{Input parameters for smaller and larger wavenumbers} \label{apsec:inputparams}

The parameters at several $\ell$ obtained
from the fitting of the observed spectra averaged over one Carrington rotation 2211 for the disc center tile 
from SDO/HMI pipeline are shown in Tables~\ref{table:input_parameters_k14} ($\ell=328$), 
\ref{table:input_parameters_k21} ($\ell=492$), and \ref{table:input_parameters_k42}  ($\ell=984$).
These average power spectra were obtained from DRMS specification 
{\tt hmi.rdvavgpspec\_fd15[2211][0][0]} and the fitting was done using the updated code.
These parameters are used as input parameters to generate Monte-Carlo synthetic data 
in Sect.~\ref{sec:performance}.

\begin{table*}[htb]
\caption{Input parameter set at $\ell=328$ ($k_\mathrm{pix}=14$)
 obtained from the fitting the average power spectrum over one Carrington rotation to the model by the updated code. See text for more details.  These parameters are used to construct the limit spectrum for Monte-Carlo simulations in Sect.~\ref{sec:performance}.
}               
\label{table:input_parameters_k14}      
\centering                                      
\begin{tabular}{rr| c c c c c c c}          
\multicolumn{9}{l}{$n$-th peak} \\
\hline
\multicolumn{2}{r|}{$q_{i,n}$ ($i=0, \dots 6$)}& $q_{0,n}$ [$\mu$Hz]& $q_{1,n}$ & $q_{2,n}$ [$\mu$Hz]& $q_{3,n}$ [m/s]& $q_{4,n}$ [m/s]& $q_{5,n}$ [$10^{-2}$]& $q_{6,n}$ [$10^{-2}$] \\
\multicolumn{2}{r|}{parameter}& $\nu_n$ & $A_n$ & $\Gamma_n$ & $u_{x,n}$ & $u_{y,n}$ & $f_{c,n}$ & $f_{s,n}$ \\
\multicolumn{2}{r|}{description} & mode frequency & amplitude & width & \multicolumn{2}{c}{horizontal velocity} & \multicolumn{2}{c}{anisotropy terms} \\
&&&&&$x$-component   & $y$-component  &  cos coeff. & sin coeff. \\
\hline
$n$&0&1825.459 &   0.613 & 118.925 &  -1.243 &  -4.119 &  -5.261 &   1.882 \\
&1&2363.872 &   4.511 &  64.812 &  -6.037 & -13.132 &  -3.101 &   3.441 \\
&2&2872.141 &  20.905 &  59.858 &   1.708 &  -5.376 &  -1.729 &   3.337 \\
&3&3341.476 &  31.187 &  76.269 &  -2.482 &  -6.713 &   9.289 &  -0.110 \\
&4&3770.138 &  13.130 &  96.789 &  -0.408 &  -4.347 &   6.617 &   7.176 \\
&5&4191.563 &   4.466 & 116.890 &   1.346 &  -3.593 &   4.760 &   7.599 \\
&6&4603.656 &   1.471 & 142.377 &  32.295 &   1.612 &   2.171 & -10.467 \\
&7&4955.544 &   0.171 & 110.552 &  13.936 & -27.390 &  37.613 &  26.478 \\
&8&5043.076 &   0.218 & 140.208 &  -2.420 &   9.940 &  -3.542 &   6.562 \\
\hline
\multicolumn{9}{l}{Background} \\
\hline
\multicolumn{2}{r|}{$q_{i,\textrm{BG}}$ ($i=0, \dots 3$)} & $q_{0,\textrm{BG}}$   & $q_{1,\textrm{BG}}$   & $q_{2,\textrm{BG}}$ [$10^{-2}$]  & $q_{3,\textrm{BG}}$ [$10^{-2}$] \\
\multicolumn{2}{r|}{parameter}  & $B_0$ & $b$ & $f_{c, \mathrm{bg}}$ & $f_{s, \mathrm{bg}}$ \\
 \multicolumn{2}{r|}{description} & amplitude & power-law& \multicolumn{2}{c}{anisotropy terms} \\
&&&index& cos coeff. & sin coeff.\\ 
  \hline
& &    1.651&   0.915&  19.670&   1.576 \\
\hline                                             
\end{tabular}
\end{table*}

\begin{table*}
\caption{Input parameters at $\ell = 492$ ($k_\mathrm{pix}=21$) similar to Table~\ref{table:input_parameters_k14}. 
}  
\label{table:input_parameters_k21}      
\centering                                      
\begin{tabular}{rr| c c c c c c c}          
\multicolumn{9}{l}{$n$-th peak} \\
\hline
\multicolumn{2}{r|}{$q_{i,n}$ ($i=0, \dots 6$)}& $q_{0,n}$ [$\mu$Hz]& $q_{1,n}$ & $q_{2,n}$ [$\mu$Hz]& $q_{3,n}$ [m/s]& $q_{4,n}$ [m/s]& $q_{5,n}$ [$10^{-2}$]& $q_{6,n}$ [$10^{-2}$] \\
\multicolumn{2}{r|}{parameter} & $\nu_n$ & $A_n$ & $\Gamma_n$ & $u_{x,n}$ & $u_{y,n}$ & $f_{c,n}$ & $f_{s,n}$ \\
\multicolumn{2}{r|}{description} & mode frequency & amplitude & width & \multicolumn{2}{c}{horizontal velocity} & \multicolumn{2}{c}{anisotropy terms} \\
&&&&&$x$-component   & $y$-component  &  cos coeff. & sin coeff. \\
\hline
$n$ & 0& 2220.729 &   1.751 &  87.719 &  -4.713 &  -1.765 &   0.066 &   0.481 \\
\multicolumn{2}{r|}{1}&2777.597 &  10.390 &  60.064 &  -2.834 &  -1.923 &   0.796 &   0.305 \\
\multicolumn{2}{r|}{2}&3337.284 &  22.333 &  56.491 &  -1.109 &  -3.998 &   0.889 &   0.769 \\
\multicolumn{2}{r|}{3}&3897.292 &  10.165 &  70.522 &   0.358 &  -2.738 &   0.962 &   0.945 \\
\multicolumn{2}{r|}{4}&4448.653 &   2.688 & 100.490 &   6.971 &  -8.530 &   1.346 &   1.261 \\
\multicolumn{2}{r|}{5}&4963.828 &   0.617 & 142.046 &   8.090 & -16.799 &   1.207 &   1.899 \\
\multicolumn{2}{r|}{6}&5465.279 &   0.111 & 158.345 &  51.843 &  -2.880 &  -0.084 &   1.761 \\
\multicolumn{2}{r|}{7}&5995.786 &   0.022 & 115.359 &  30.740 &   5.315 &  -2.908 &   1.227 \\
\hline
\multicolumn{9}{l}{Background}\\
\hline 
\multicolumn{2}{r|}{$q_{i,\textrm{BG}}$ ($i=0, \dots 3$)} & $q_{0,\textrm{BG}}$   & $q_{1,\textrm{BG}}$   & $q_{2,\textrm{BG}}$ [$10^{-2}$]  & $q_{3,\textrm{BG}}$ [$10^{-2}$] \\
\multicolumn{2}{r|}{parameter}  & $B_0$ & $b$ & $f_{c, \mathrm{bg}}$ & $f_{s,\mathrm{bg}}$ \\
 \multicolumn{2}{r|}{description} & amplitude & power-law & \multicolumn{2}{c}{anisotropy terms} \\
&&&index& cos coeff. & sin coeff.\\ 
  \hline
 &&    1.959&   0.947&   4.484&  -0.059 \\
\hline                                             
\end{tabular}
\end{table*}

\begin{table*}[htb]
\caption{Input parameter set at {$\ell=984$ ($k_\mathrm{pix}=42$)}, similar to Table~\ref{table:input_parameters_k14}. }               
\label{table:input_parameters_k42}      
\centering                                      
\begin{tabular}{rr|ccccccc}          
\multicolumn{9}{l}{$n$-th peak} \\
\hline
\multicolumn{2}{r|}{$q_{i,n}$ ($i=0, \dots 6$)}& $q_{0,n}$ [$\mu$Hz]& $q_{1,n}$ & $q_{2,n}$ [$\mu$Hz]& $q_{3,n}$ [m/s]& $q_{4,n}$ [m/s]& $q_{5,n}$ [$10^{-2}$]& $q_{6,n}$ [$10^{-2}$] \\
\multicolumn{2}{r|}{parameter}& $\nu_n$ & $A_n$ & $\Gamma_n$ & $u_{x,n}$ & $u_{y,n}$ & $f_{c,n}$ & $f_{s,n}$ \\
\multicolumn{2}{r|}{description} & mode frequency & amplitude & width & \multicolumn{2}{c}{horizontal velocity} & \multicolumn{2}{c}{anisotropy terms} \\
&&&&&$x$-component   & $y$-component  &  cos coeff. & sin coeff. \\
\hline
$n$&0&3100.633 &   5.476 & 156.240 &  -7.238 &  -4.020 &   0.802 &   1.543 \\
&1&3832.691 &   5.273 &  84.944 &  -9.163 &  -2.148 &   1.640 &   1.493 \\
&2&4560.268 &   1.703 & 122.403 &  -3.386 &  -3.871 &   1.317 &   1.712 \\
&3&5291.354 &   0.331 & 253.910 &  10.232 &  -3.632 &   1.270 &   1.150 \\
\hline 
\multicolumn{9}{l}{Background} \\
\hline
\multicolumn{2}{r|}{$q_{i,\textrm{BG}}$ ($i=0, \dots 3$)} & $q_{0,\textrm{BG}}$   & $q_{1,\textrm{BG}}$   & $q_{2,\textrm{BG}}$ [$10^{-2}$]  & $q_{3,\textrm{BG}}$ [$10^{-2}$] \\
\multicolumn{2}{r|}{parameter}  & $B_0$ & $b$ & $f_{c, \mathrm{bg}}$ & $f_{s, \mathrm{bg}}$ \\
 \multicolumn{2}{r|}{description} & amplitude & power-law & \multicolumn{2}{c}{anisotropy terms} \\
&&&index& cos coeff. & sin coeff.\\ 
  \hline
& & 1134.940&   1.897&   5.735&   0.818 \\
\hline                                             
\end{tabular}
\end{table*}

\section{Underestimated amplitude -- Logarithm of average and average of logarithm} \label{apsec:app_amp}

Figure~\ref{fig:kr21_p_dp} shows that
the amplitudes ($q_{1,n}$ for all $n$ and $q_{0,b}$ in Eq.~(\ref{eq:altered_Pmodel}))  
measured by the updated code are underestimated.
This is the result of taking the logarithm of the rebinned power and 
the effect can be reproduced in the following simple test.

To see why this is the case, we define $x_i \ (i =0, 1, \dots n-1)$ as a series of random variables whose distribution function is the chi-square distribution with two degrees of freedom.
The expectation values and the variance of $x$ are 
$\langle x \rangle =2$ and $\sigma_x^2= 2^2$, respectively,
and $y_j  (j =0, 1, \dots n/a-1)$ is smoothed $x_i$ over each $a$-pixel range:
\begin{eqnarray}
 y_j = \frac{1}{a}\sum_{i=j a}^{(j+1)a-1} x_i \  .
\end{eqnarray}
In this case $\langle y \rangle = \langle x \rangle = 2$, and $\ln\langle y \rangle= \ln(2)$,
but $\langle \ln (y) \rangle \le \ln\langle y \rangle $. The more the array is rebinned (the larger $a$ becomes), the less the scatter of $y$ becomes: $\langle \ln(y)\rangle $ approaches $\ln\langle y \rangle $.

\begin{figure}[hbtp]
\centering
\includegraphics[width=0.48\textwidth]{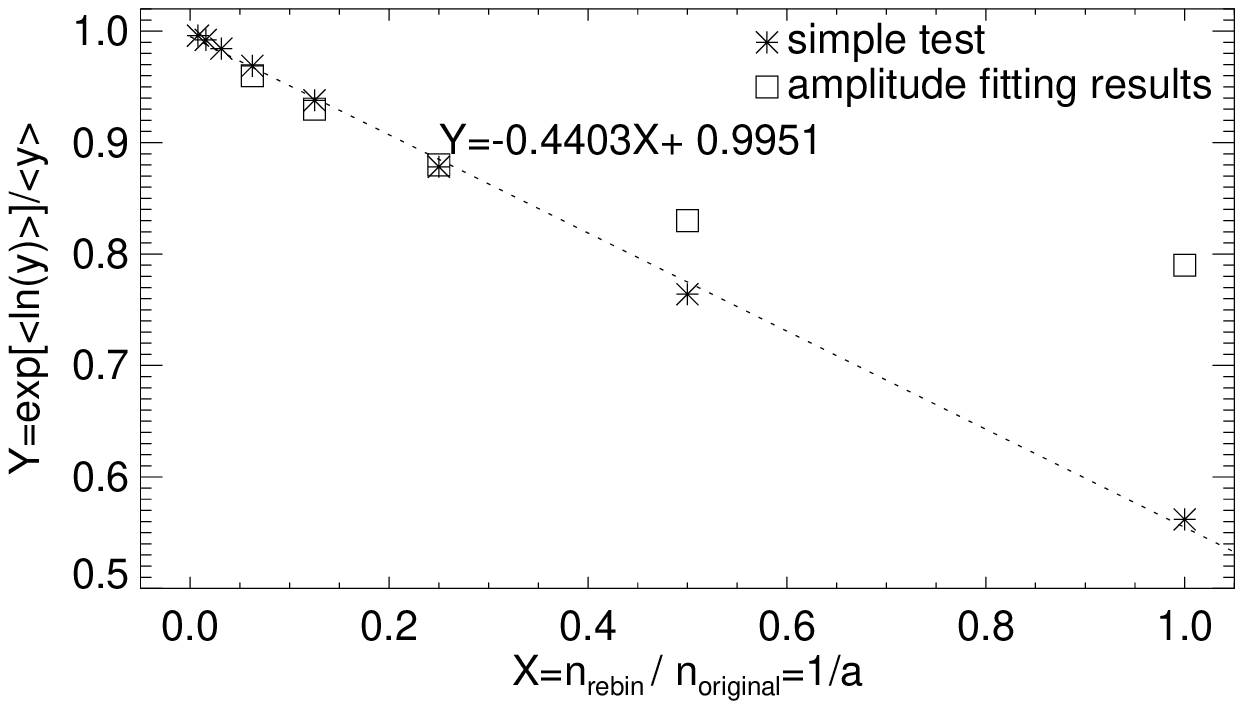} 
\caption{$\exp{\langle\ln (y)\rangle} / \langle y \rangle$, where $y$ is the
$a$-element average of chi-square distributed random variables, 
as a function of the pixel number ratio of the rebinned data to the original (asterisks). The average ratio of the amplitude fitting results to the input shown in Fig.~\ref{fig:size_aratio} is depicted by squares. The dashed line and the equation on the panel are the linear regression of the data points.}
\label{fig:20190312_test_Exln_lnEx}
\end{figure}

\begin{figure}[htbp]
\centering
\includegraphics[width=0.24\textwidth]{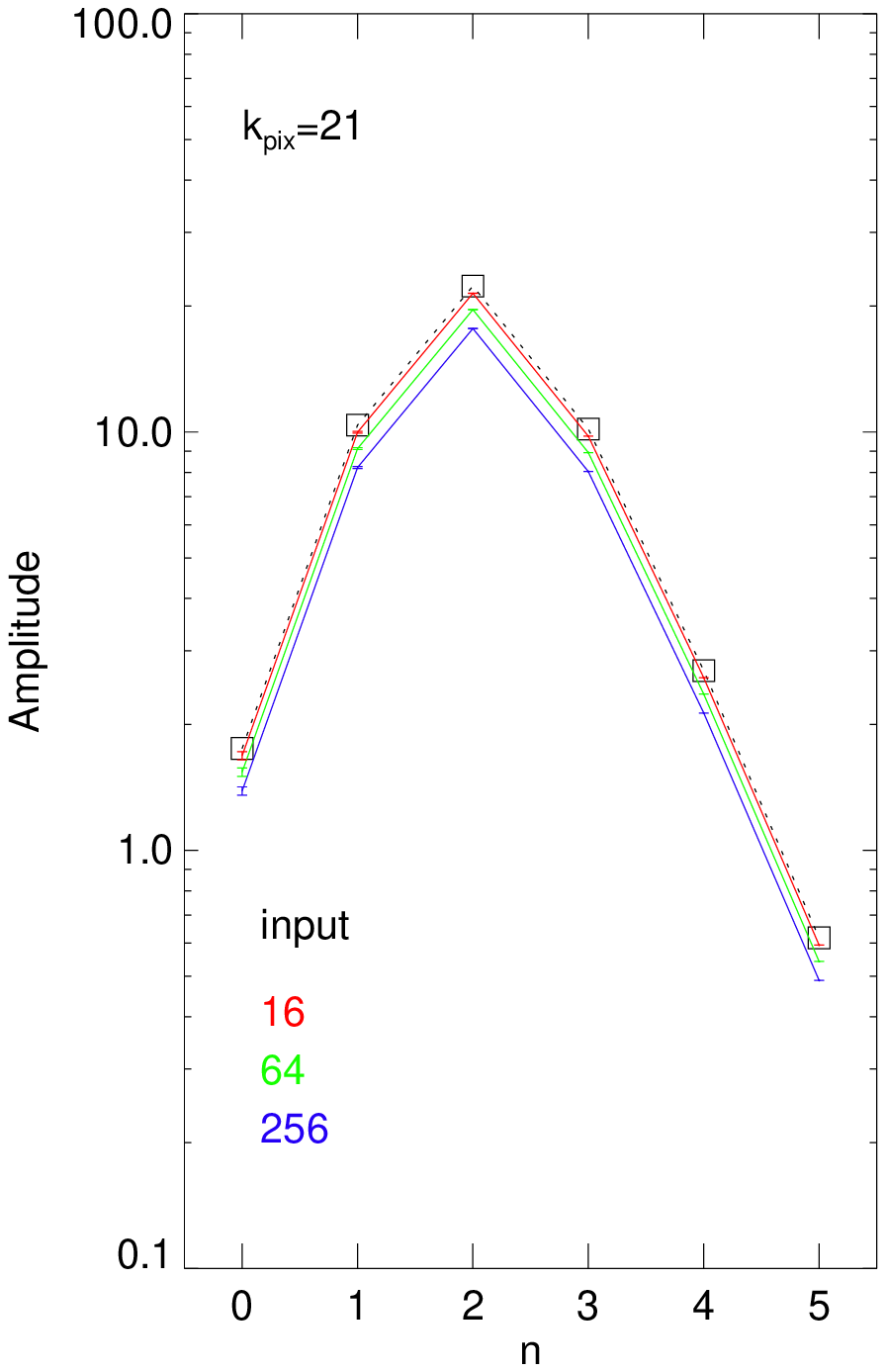}
\includegraphics[width=0.24\textwidth]{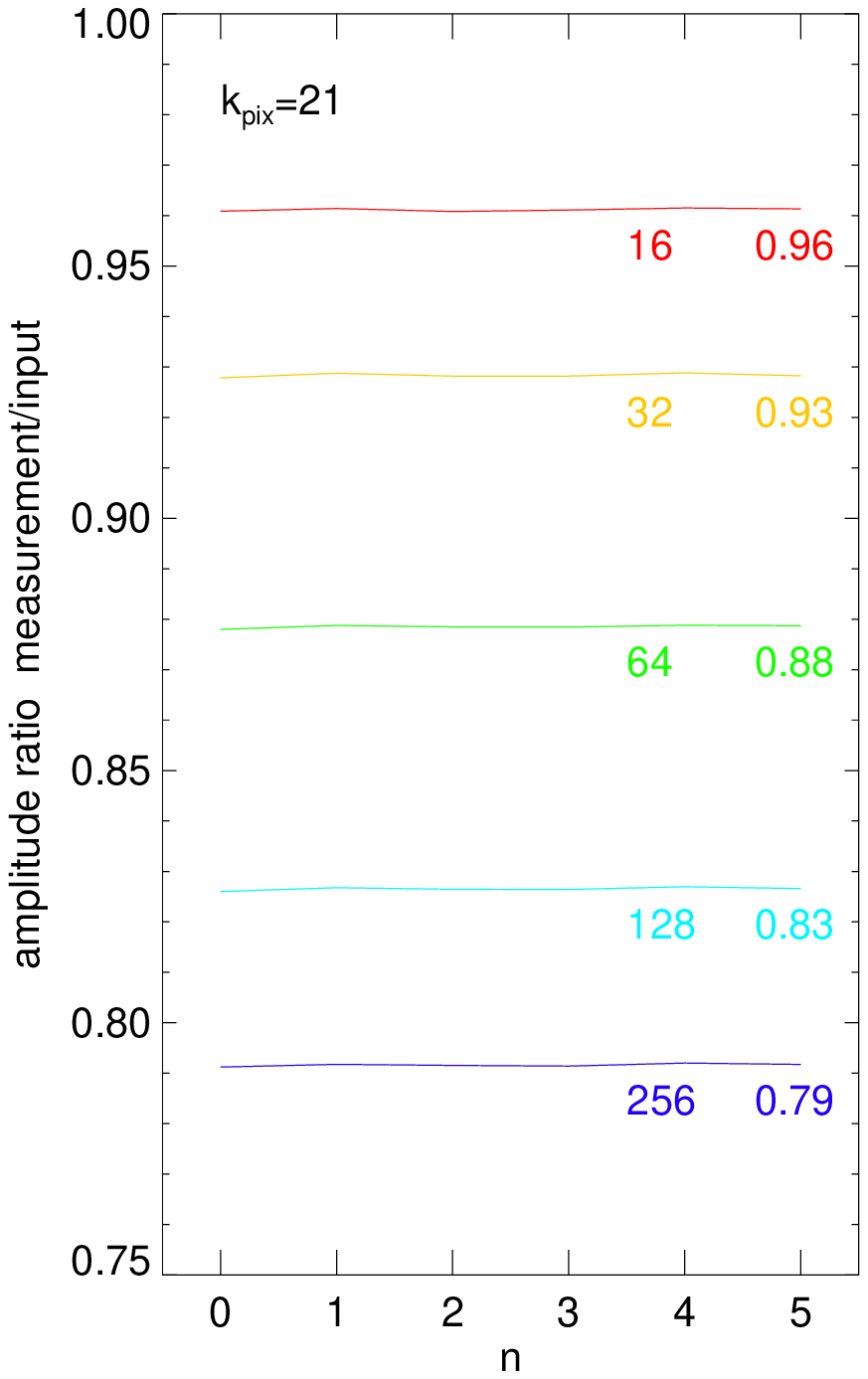} 
\caption{Dependence of the peak amplitude parameters at $\ell = 492$ ($k_{\mathrm{pix}} =21$) measured by the updated code on the rebinning strength. Left panel shows the input (black dotted lines with squares) and measurements with three different rebinning with error bars (scatter of 500~realizations): no rebinning (256~pixels on the azimuthal grid, blue), original rebinning (64~pixels, green), and extra rebinning (16~pixels, red). The right panel shows the ratio of the amplitude to the input. The numbers are the grid number and the average ratio of six peaks. The input parameters are identical to those in Fig.~\ref{fig:kr21_p_dp}.}
\label{fig:size_aratio} 
\end{figure}

This trend is alleviated if the spectra are more strongly averaged.
Figure \ref{fig:20190312_test_Exln_lnEx} shows a simple test calculation to illustrate how the average of logarithm is reduced from the logarithm of the average;
the ratio of $\langle \ln (y) \rangle $ to $ \ln\langle y \rangle $ is plotted against the pixel number ratio of the rebinned data to the original. 
In this simple test calculation, we make 500 realizations of sets of $x_i \  (i =0, 1, \dots n-1, \mathrm{where} \  n=256)$, calculate $y$ with $a=1, 2, 4, 8, \dots 256 $, and 
compute how much smaller the expectation value of the logarithm of $y$ than the logarithm of the expectation value is. The linear regression of the data points are also shown in the plot. For comparison the amplitude fitting results shown in Fig.~\ref{fig:size_aratio} are replotted here with squares.
This explains how the logarithm of amplitude is reduced.
In the case of no rebinning ($X=n_{\rm rebin}/ n_{\rm original} =1 $) or only 2-pixel rebin ($X=1/2$) the fitting results are deviated significantly from this simple test calculation, but this comes as no surprise because in these cases, the assumption of the well-averaged power as data is not appropriate.

Figure~\ref{fig:size_aratio} shows the dependence of the amplitude measurements on the amount of rebinning. 
The input model is identical to the one used in  the calculation in Sect.~\ref{subsec:error_estimate}.
The default rebinning is from 256~pixels 
to 64~pixels and in this case, the output amplitude is about 88\% of the input amplitude but with further rebinning to 16~pixels, it increases up to 96\%. 
 
\end{document}